\documentclass[12pt]{article}
\pdfoutput=1

\usepackage{aas_macros,amsmath,amssymb,cite,esint,graphicx,mathtools,transparent}

\usepackage[margin=1in,a4paper]{geometry}
\usepackage[colorlinks=true]{hyperref}
\usepackage[affil-it]{authblk}
\usepackage[normalem]{ulem}
\usepackage[compact]{titlesec}  
\usepackage{grffile}

\usepackage[absolute,overlay]{textpos}

\usepackage{epsfig}
\usepackage{amsfonts}
\usepackage{latexsym}
\usepackage{amsmath}
\usepackage{mathrsfs}
\usepackage{hyperref}
\usepackage{wasysym}
\usepackage{tikz}
\usepackage{caption}
\usepackage{subcaption}
\usepackage{float}
\usepackage{comment}
\numberwithin{equation}{section}

\DeclareFontFamily{OT1}{rsfs}{}
\DeclareFontShape{OT1}{rsfs}{m}{n}{
<-7> rsfs5 <7-10> rsfs7 <10-> rsfs10}{}
\DeclareMathAlphabet{\mycal}{OT1}{rsfs}{m}{n}

\newcommand{\bea}{\begin{align}}
\newcommand{\eea}{\end{align}}

\def\ads2{$\mathrm{AdS_2}$}
\def\nads2{$\mathrm{NAdS_2}$}
\def\sl2r{$\mathrm{SL(2,R)}$}

\begin{document}

\unitlength = 1mm

\setcounter{tocdepth}{2}

\title{\textbf{\Huge {Exploring lensing signatures through spectrotemporal correlations: implications for black hole parameter estimation}}}

\date{}
\author[1,2]{Sreehari~Harikesh\thanks{hjsreehari@gmail.com}}
\author[2,3]{Shahar~Hadar\thanks{shaharhadar@sci.haifa.ac.il}}
\author[1,2]{Doron~Chelouche\thanks{doron@sci.haifa.ac.il}}

\affil[1]{\footnotesize Department of Physics, Faculty of Natural Sciences, University of Haifa, Haifa 3498838, Israel
}
\affil[2]{\footnotesize Haifa Research Center for Theoretical Physics \& Astrophysics, University of Haifa, Haifa 3498838, Israel
}
\affil[3]{\footnotesize Department of Mathematics and Physics, University of Haifa at Oranim, Kiryat Tivon 3600600, Israel
}

\maketitle


\begin{abstract}
Extreme gravitational lensing and relativistic frequency shifts, combined together, imply that radiation emitted from a black hole's vicinity can echo at different frequencies and times, leading to spectrotemporal correlations in observed signals.
If such correlations are uncovered by future observations, they could provide a probe of the spacetime geometry in the strong-field region near black holes.
Here, motivated by these prospects,
we numerically compute the two-point correlation function of specific flux fluctuations in a simple model of line emission by a hotspot in an equatorial circular orbit. We make use of the Adaptive Analytical Ray Tracing (AART) code to generate the light curves we then correlate. Our results for the correlation maps show a clear decomposition into direct emission-dominated, and lensing-dominated contributions. The computation transcends past analytical approximations, studying the main contribution to the correlation function, which is not deep in the universal regime. We compute correlation maps for many combinations of black hole mass, spin, inclination, hotspot width, and orbital radius, and study their dependence on these parameters. The correlation maps are then used to train convolutional neural networks which can be used to estimate source parameters, achieving promisingly low evaluation errors within the model. Our results could be relevant for future X-ray spectroscopic missions, 
offering insights into black hole parameter inference.\\

(\textcolor{blue}{Received 25 February 2025; accepted 22 July 2025} in Physical Review D)\\
\end{abstract}

\pagestyle{plain}
\setcounter{page}{1}
\newcounter{bean}
\baselineskip18pt


\setcounter{tocdepth}{2}

\section{Introduction}\label{sec:introduction}
Observed electromagnetic radiation from the environs of a black hole (BH) 
is determined by the energetic astrophysical phenomena that generate it and the strong gravitational field that lenses it. 
In BH images, extreme gravitational lensing of light rays is expected to result in a photon ring \cite{Teo2003,Beckwith2005,Johannsen2013,Gralla2019}.
The photon ring is the collection of all images of the source which are extremely lensed, i.e. have large deflection angles, and undergo multiple half-orbits $n\geq1$ around the BH, where we define an orbit as a full oscillation in the polar angle with respect to the BH spin axis; in contrast, the direct image is the contribution of photons which undergo $n=0$ half-orbits around the BH before reaching a far observer.
The importance of the photon ring is in its universality: while the direct image depends strongly on complex source astrophysics, the photon ring is a prediction of general relativity (GR). It has simple characteristics \cite{Johnson2020,Gralla2020lensing} that depend on the spacetime geometry and therefore could allow future tests of GR, and serve as a means to deduce the properties of astrophysical BHs.  

However, despite the remarkable recent progress by the Event Horizon Telescope (EHT) \cite{EHT2019a,EHT2019d,EHT2022a,EHT2022b,EHT2022c,EHT2022d,EHT2022e}, photon rings have not yet been detected in interferometric images. Significant efforts are underway to plan new missions that could allow interferometric detection of the photon ring.
The next-generation Event Horizon Telescope (ngEHT) plans to extend EHT by adding several dedicated multifrequency radio telescopes around the globe, significantly improving visibility-space coverage and temporal monitoring capabilities. 
The Black Hole Explorer (BHEX) \cite{Johnson2024} is set to overcome terrestrial constraints by extending sub-millimeter Very-Long-Baseline Interferometry (VLBI) into space, opening the door to an order-of-magnitude increase in angular resolution that should guarantee a detection of the photon ring in M87* or SgrA*, assuming environmental effects do not obscure the ring in both of these targets.

In parallel to the development of such groundbreaking instruments, there is high demand for novel techniques that disentangle the effects of strong gravity from astrophysical signals.
One proposal that this has led to is an alternative method for detecting photon rings, presented in~\cite{Hadar2021}. The observable studied there is the two-point spatial (auto-)correlation function of fluctuations in intensity along the photon ring. 
Such correlations must arise since any source fluctuation near a BH has multiple (formally, an infinite number of) images at relative locations and times dictated by the spacetime geometry. Therefore, even a noisy incoherent source will have a noisy image but not a completely incoherent one, as it will contain hidden correlations.
This method has some potential advantages to time-averaged imaging: for example, it could be applicable when dealing with observations of rings in which the diameter is resolved but the photon ring width is not. Instead, temporal resolution at timescales shorter than the characteristic BH timescale is required, among other conditions.
In \cite{Zhang2025}, the authors considered autocorrelations of lensed emission in BH images in the context of an orbiting hotspot source, and demonstrated that this approach could be used to disentangle lensing effects from intrinsic source correlation.
See also \cite{Wong2021,Chesler2021} for related approaches to photon ring temporal signatures,
\cite{Cardenas-Avendano2024} for an explanation why single-frequency total light curves are insufficient for resolving lensing correlations in the radio, 
and \cite{Wong2024} for a related visibility-space correlation observable, especially relevant for BHEX.
Recently, we \cite{Hadar2023} have proposed a generalization of this observable: the spectrotemporal (auto-)correlation function (STAC). Of particular interest in the present context is the image-integrated version of the STAC observable, which is relevant for sources that are totally unresolved spatially but are resolved spectrally and temporally. This observable exchanges the image positions, which are the relevant parameters in Ref.~\cite{Hadar2021}, with the observed frequencies of the detected photons\cite{Hadar2023}. A disadvantage is that one must make assumptions on the source in order to translate between image position and relativistic red/blueshift; an advantage is that this method is potentially applicable to many sources, including at high redshifts.

X-ray observations provide a powerful method for studying relativistic effects near BHs, particularly through the broadening and skewing of spectral lines emitted by ions in the accretion disk \cite{Laor1991,Fabian2000,Nandra2001,Reynolds2003,Seward2010}. These lines, which are associated with fluorescence from the accretion flow, become broadened due to GR effects, namely Doppler shift and gravitational redshift, on photons emitted close to the BH. As the gas in the inner regions of the accretion disk moves at velocities approaching the speed of light, the emitted X-rays are distorted, providing a signature of the strong gravitational field near the BH; for recent semi-analytical treatments see \cite{Gates2020,Gates2024}.
One of the key techniques used to analyze these relativistic effects is X-ray reverberation mapping \cite{Reynolds1999,Zoghbi2012,Uttley2014}. In this method, X-rays emitted from the central corona illuminate the accretion disk, and the reflection of these X-rays can be detected with a time delay. The lag in the reflection signal allows us to estimate the properties of the source.
In addition, the study of time lags between different energy bands \cite{Wilkins2016,Kara2016}, such as soft and hard X-rays, has revealed important information about the structure of the accretion flow.
In recent years, advances in X-ray observatories have significantly improved our ability to probe these relativistic effects. High-resolution observations from missions like XMM-Newton \cite{Jansen2001}, NuSTAR \cite{Harrison2013}, and NICER \cite{Gendreau2016} have allowed more detailed studies of X-ray spectra from both active galactic nuclei (AGN) and X-ray binaries (XRBs). Future telescopes, such as the NewAthena mission \cite{Athena_10FIRST_2013}, are expected to push these capabilities further, providing even finer spectral resolution and enhanced sensitivity.

In \cite{Hadar2023}, we performed analytical calculations in a `photon ring' approximation, which assumes that $n\gg1$\footnote{In practice, this approximation converges quite rapidly: $n\geq2$ contributions are deep in its regime of validity, and it often describes even $n=1$ contributions reasonably well.\label{footnote:large n}}, and showed that STACs are universally manifested as a result of extreme lensing by BHs. However, the leading (flux-wise) contributions to STAC arise from autocorrelation of the $n=0$ contribution---which is the direct signal, which is not expected to probe extreme lensing effects---and from correlation of the $n=0$ and $n=1$ contributions. Even if we take into account its rapid convergence (see footnote \ref{footnote:large n}), these leading contributions are not under control of the large-$n$ approximation. It is therefore very interesting to compute them numerically and understand how their details deviate from simple large-$n$ analytical predictions.
In this paper, 
we numerically compute these leading STACs for a simplified source model. We assume a hotspot on a circular geodesic orbit around a Kerr BH, emitting monochromatically in its rest frame. Relativistic effects blue/redshift the emitted photons and result in STACs in the signals we observe, which we calculate. The idea is to provide a simple toy model for fluctuations in, say, the $\sim$6.4 keV K$\alpha$ iron-line emission in the close vicinity of a BH, within which an identification of extreme lensing signatures, and parameter inference, would be especially tractable.
As a key tool in our study, we use a new ray-tracing code which is tailored for computations of extreme lensing effects: the adaptive analytical ray tracing (AART) code \cite{Alejandro2023}. 
In AART, ray tracing is performed with a resolution which is adapted per half-orbit number $n$. Namely, it must grow exponentially with $n$ in order to faithfully trace photon-ring images. Moreover, AART optimizes ray tracing using closed-form analytical formulae \cite{Gralla2020null_geodesics}. Note that ray-tracing methods with a multiscale approach have also been implemented in \cite{Gelles2021} to generate realistic images of BH sources using the IPOLE \cite{Moscibrodzka2017} code.

Using the series of images---or `movies'---that we compute using AART, we can infer the blue/redshift of the observed photons and produce light curves corresponding to each observation frequency. We then compute the correlations of fluctuations in these light curves to produce two-point correlation functions. 
Once we generate these correlation maps, we apply machine learning algorithms---namely,  convolutional neural networks (CNNs)---to estimate the spin, inclination, and mass of any given BH source. Additionally, we estimate the location of the monochromatic hotspot source around the BH. 

Prior to this, the authors of \cite{vander2020} have applied CNNs to estimate the mass, spin, and accretion rate of the BH source based on 230 GHz and 690 GHz data. Recently, \cite{Farah2024} implemented a CNN based on the bayesian Deep Horizon neural network by \cite{vander2020} which was introduced primarily for BH parameter estimation. A neural network trained to estimate the inclination of a BH was generated as presented in reference \cite{Popov2021} based on interferometric images. In contrast, our algorithm will work with X-ray data as input. The method we propose here can be applied on both X-ray binaries and supermassive BHs and is independent of BH parameters, though spin effects will be more evident for relatively fast spinning sources having moderately high inclinations. 

The paper is organized as follows. We explain how we extract energy-specific light curves using the AART code in \S \ref{sec:AART} for an emissivity profile corresponding to a monochromatic hotspot. Our results for the correlation maps, including a discussion of lensing signatures and dependence on BH parameters, are presented in \S \ref{sec:results}. We apply Machine Learning algorithms to extract parameters such as spin, inclination, mass and hotspot location of the source from these data in \S \ref{sec:ML}. Subsequently, we mention some observational considerations in \S \ref{sec:ForObs} and conclude in \S \ref{sec:disc}.
We work with geometric units $G=c=1$.

\section{Methodology and source model}
\label{sec:AART}

\begin{figure}
    \centering
    \includegraphics[width=0.45\linewidth]{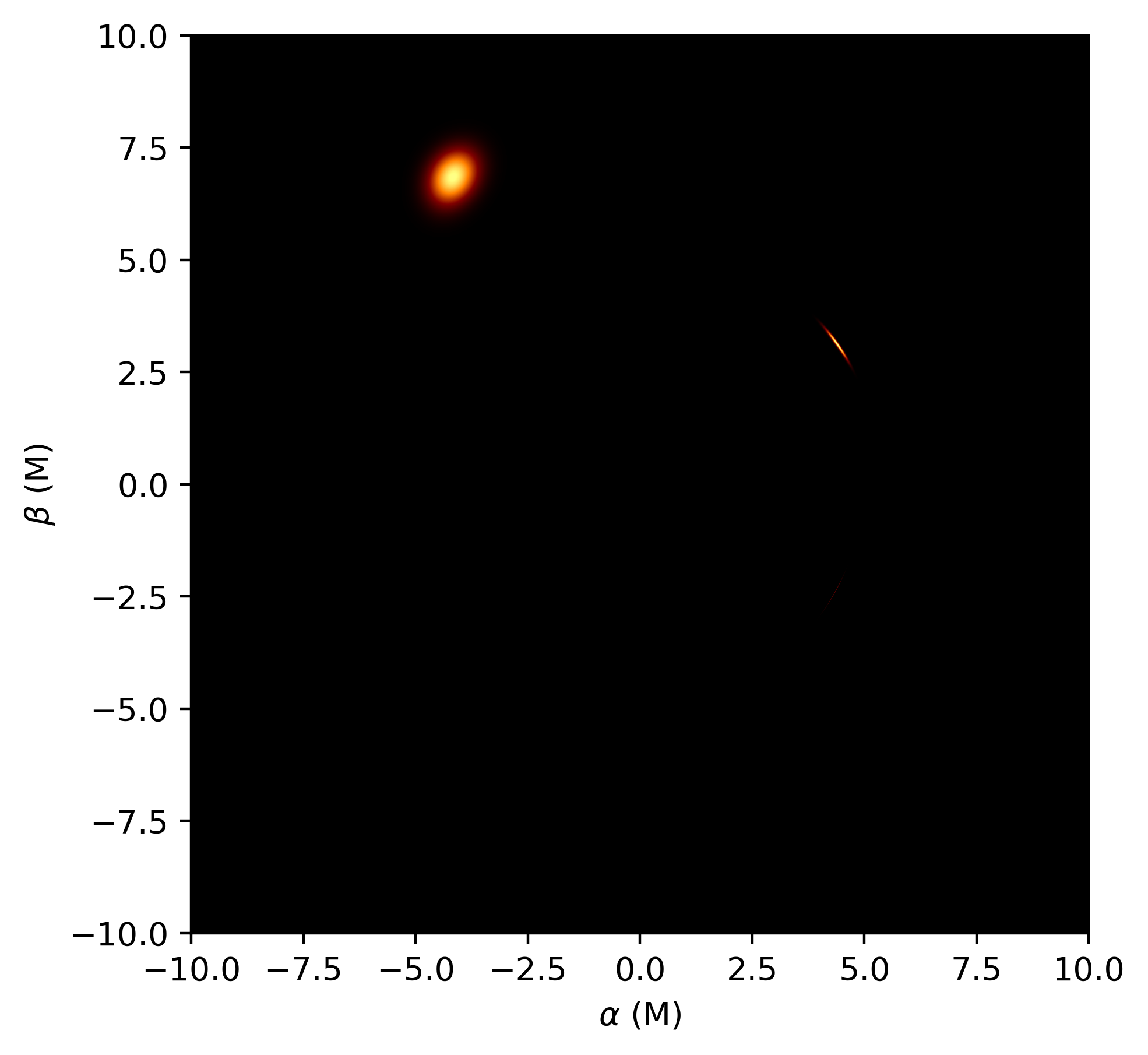}
    \includegraphics[width=0.45\linewidth]{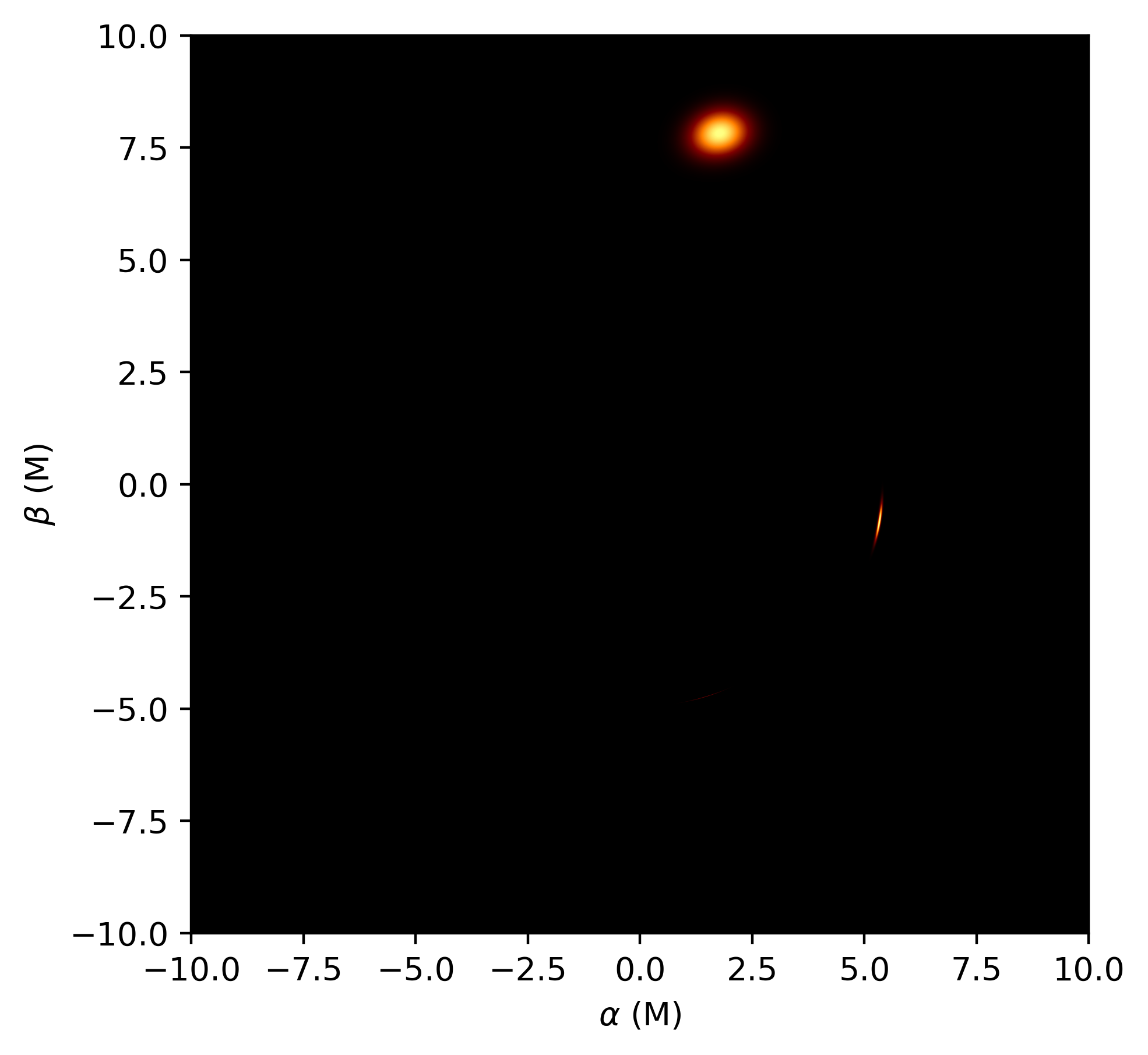}
    \caption{Snapshots of a hotspot orbiting a Kerr black hole of dimensionless spin $a=0.9$, viewed at an inclination of 0.5$^\circ$, generated using the AART code \cite{Alejandro2023}. The black hole is aligned with the center of the image. The hotspot orbits along a circular equatorial geodesic at a radius $r_s=7M$ with angular velocity $\Omega \approx 0.05 M^{-1}$. The two snapshots are separated by a time $14M$. 
    Along with the direct $n=0$ image, a demagnified $n=1$ image and a faint $n=2$ image, arising from extremely lensed light rays, are noticable. $\alpha,\beta$ are impact parameters of observed light rays. See text for details.}
    
    \label{fig:hotspot}
\end{figure}

In this section, we describe our geometric toy model for the radiation source and the techniques we used in order to numerically compute the spectrotemporal correlations of the fluctuations of its specific flux. Our model consists of a hotspot orbiting the BH along a corotating geodesic equatorial circular trajectory which is isotropically emitting monochromatic light (for concreteness, we take line emission from low-ionization iron at 6.4\,keV \cite{George1991,Turner2002}) in its rest frame. Fig.~\ref{fig:hotspot} shows an AART-generated \cite{Alejandro2023} snapshot of such a source, orbiting a Kerr BH of mass $M$ and dimensionless spin $a=J/M^2=0.9$, where $J$ is the BH angular momentum, taken at a Boyer-Lindquist source radius $r_s=7M$, angular velocity $\Omega = \frac{\pm\sqrt{M}}{r_s^{3/2}\pm a\sqrt{M}}\approx 0.05 M^{-1}$ where the $\pm$ signs correspond to co-/counter- rotating orbits, respectively, and width $W=\sqrt{2}\sigma=0.5 M$, where $\sigma$ is the standard deviation of a 2D Gaussian representing the hotspot's local emissivity profile. In this figure, the images corresponding to $n=1$ and $n=2$ (very faint) in each panel are exponentially demagnified versions of the main emission. In an optically thin scenario, the surface brightness does not significantly decrease with $n$, while the flux decreases exponentially with $n$ due to this demagnification.

Relatedly, an important comment is in order: we assume that at least parts of the disk in the region where extreme lensing takes place---called the photon shell \cite{Johnson2020}---are transparent enough to allow light rays to orbit the BH. One may imagine various situations which could lead to such a scenario. For example, if the flow is clumpy and displays significant density inhomogeneities, localized pockets of neutral iron emission could form within a predominantly Compton-thin gas; see \cite{Dexter2011} for discussion of such models. Another possibility is that in the region around the innermost stable circular orbit (ISCO) or below it, the density drops to a low enough value to allow light rays to orbit, at least in part of the photon shell. Alternatively, the accretion flow may be truncated, perhaps due to outflows and the opening of gaps \cite{Jac2021}, or assume a puffed-up optically thin inner corona configuration \cite{Rapisarda2016,Noda2018}.

Using the AART code \cite{Alejandro2023}, we have traced the light rays that come directly to the observer ($n=0$), as well as the rays that undergo one ($n=1$) and two ($n=2$) half-orbits around the BH before reaching the observer. 
The output intensity for any $n$ is a product of the input emissivity profile and the redshift factor cubed, $g^3$. Here, using Cunningham's analytic results \cite{Cunningham1975}, $g$ is computed separately for sources above the ISCO (assumed to be circular orbiters) and below the ISCO (assumed to be plunging on trajectories starting approximately from the ISCO). See, for example, Eqn.~B13 of Ref.~\cite{Alejandro2023} for the relevant expressions.
The intensity and the redshift factor for each of the 512 input snapshots generated at a time resolution of 0.25 M are the outputs of the \texttt{slow\_light} or  \texttt{flare\_model} function in AART.
We obtain the energies corresponding to line emission of 6.4 keV by multiplying the line energy by the redshift factor ($g$). From this distribution of energies we can bin photons in the desired energy ranges and thereby produce light curves. 

Thus, ensuring that we span the possible redshifts/blueshifts of the iron line, light curves were generated in the range of 0.2 keV to 11.2 keV in 220 energy bins each of size 0.05 keV. Separate light curves are generated for different values of the half-orbit number $n$, making use of corresponding images for each $n$ produced with the AART code. 
In Fig. \ref{fig:LC_hotspot} we show examples of light curves that were generated with AART for a hotspot (at ISCO) orbiting a BH which has a spin of 0.8 and inclination of 40$^{\circ}$. The light curves of the main emission ($n=0$) and the lensed emissions are portrayed row-wise. The first column shows the total light curve in the energy range 0.2$-$11.2 keV and the second and third columns represent light curves at 3.2 keV and 5.2 keV respectively. The orbital period ($\sim 40$ M) of the geodesic hotspot is evident from the light curves.
Following \cite{Hadar2023}, we then computed the cross correlations between these light curves for all possible combinations of $n \in \{0, 1, 2\}$ with \texttt{stingray} software \cite{Huppenkothen2019}. The correlation matrices $C_{\nu \nu'}(T)=\langle \Delta f_{\nu}(t) \Delta f_{\nu '}(t+T) \rangle$, where $\Delta f_\nu(t)$ are the specific flux fluctuations, are generated for numerous different values of BH spin and inclination. Let $n_1,n_2$ denote the half-orbit numbers of the first and second light curves, respectively, and define the difference $m = n_1 - n_2$ which takes integer values from $-2$ to $2$.
In order to understand the effect of extreme lensing on spectrotemporal correlation maps, we find it instructive to present the $m=0$ (generically dominated by $n_1=n_2=0$) and $m\neq0$ contributions (generically dominated by $n_1=0 , n_2=1$ and $n_1=1 , n_2=0$) separately before combining them in a third total version including all-$m$ contributions. This is how correlation maps are presented throughout the paper.   
It is important to note that the correlation maps $C_{\nu \nu'}(T)$ are three-dimensional (3D), depending on two frequencies $\nu,\nu'$ and a time lag $T$. For simplicity and for clarity of presentation, we often focus on particular 2D slices of the total correlation map.
We consider here two such 2D slices. The first is defined by fixing $\nu^{\prime} = \nu$, so that the light curves corresponding to $n_1$ and $n_2$ belong to the same frequency bin. The second is defined by fixing one of the frequencies to be $\nu^{\prime}=\Tilde{g}_0\nu_{line}$ which corresponds to a line emission of 6.4 keV redshifted by the factor $\Tilde{g}_0$ \cite{Hadar2023}, corresponding to the redshift of the zero-azimuthal angular momentum critical photon orbit.

For a geodesic equatorial hotspot in a circular orbit at $r_s=7M$, with an angular velocity of 0.05 $M^{-1}$, for example, one finds correlation ridges at a time delay corresponding to the orbital period of 2$\pi/\Omega = 122M$.   
We have checked that if we artificially increase the orbital velocity of the hotspot (deviating from geodesic motion), the correlation ridges will appear at lower time delays, as expected. As the orbital velocity increases, the overlap time of the hotspot with itself in successive periods will reduce, and therefore the correlation ridge will become narrower along the time axis. However, for further analysis we consider only geodesic equatorial hotspots. 

We generate correlation patterns from light curves corresponding to the hotspot input for various combinations of spin, inclination, and hotspot size. Initially, the hotspots were placed so that their lower edge coincides with the ISCO radius, which is a function of the BH spin. The size of the hotspot was randomly chosen to be a value between 0.1 M and 0.5 M.
However, since the period of an ISCO orbiter is a rather sensitive function of BH spin, fixing the hotspot radius at the ISCO introduces a (probably) unrealistic direct dependence of the correlation patterns on source spin.
Therefore, in order that our toy model makes a significant step towards realism, we have then introduced a random hotspot location, where the hotspot orbital Boyer-Lindquist radius $r_s$ is a random value between the ISCO and 10~M.
This is done keeping in mind that the correlation patterns will eventually be used to train a machine learning model that could predict spin and inclination of the source from observed data. 

\section{Results}
\label{sec:results}
In our previous work \cite{Hadar2023}, we analytically computed correlation patterns corresponding to stochastic disc emission. That computation was enabled by assuming that the relevant light rays are near-critical, or formally that $n \gg 1$. Moreover, those analytic results assume a particularly transparent form only for small inclinations, or equivalently a narrow energy band.
Therefore, in order to accurately account for the leading contributions to the spectrotemporal correlations ($n_1=n_2=0$ for direct emission-dominated correlations and $n_1=0$, $n_2=1$ or vice versa for lensing-dominated correlations), and in order to obtain clearer heuristics of the behavior of correlations for all possible source inclinations, we rely on numerical computations in this paper. 
We emphasize here some key aspects of our findings:

\begin{figure*}
    \includegraphics[width=\textwidth]{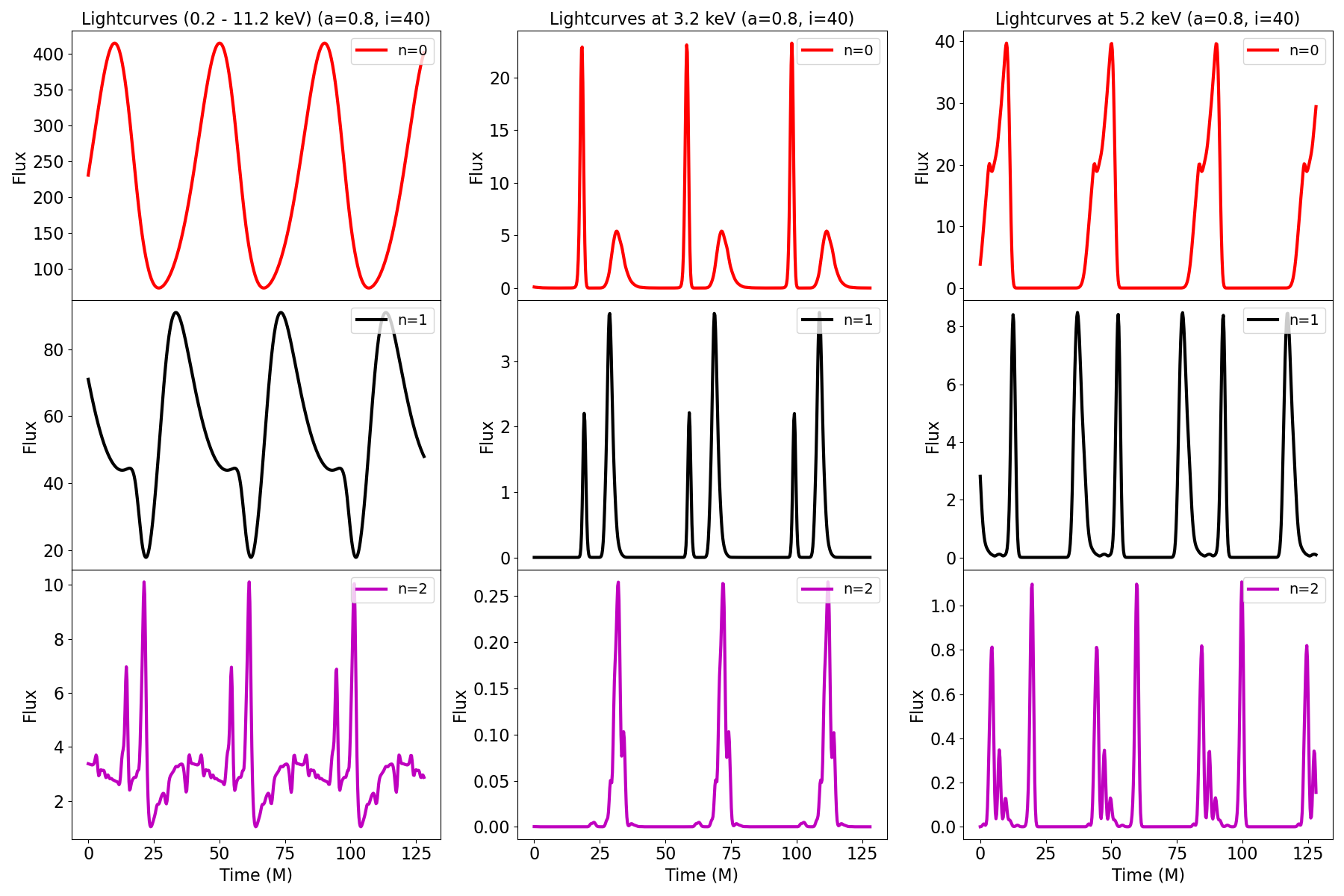}
    \caption{Example light curves generated using the procedure outlined in \S \ref{sec:AART}. The first column shows the total (summed over energy) light curves.
    The second and third columns show the 3.2 keV and 5.2 keV light curves, respectively, where each bin is of width 50 eV. The flux is presented in arbitrary units, and time is measured in units of the black hole mass $M$. The three rows correspond to $n=0$, 1, and 2 respectively. All panels in this figure correspond to a source with dimensionless spin 0.8 and inclination 40$^{\circ}$ with an orbiting hotspot at the innermost stable circular orbit.}
    \label{fig:LC_hotspot}
\end{figure*}

\subsection{Identifying signatures of extreme lensing}
\label{sec:lensing}

\begin{figure}
    \centering
    \includegraphics[width=1\linewidth]{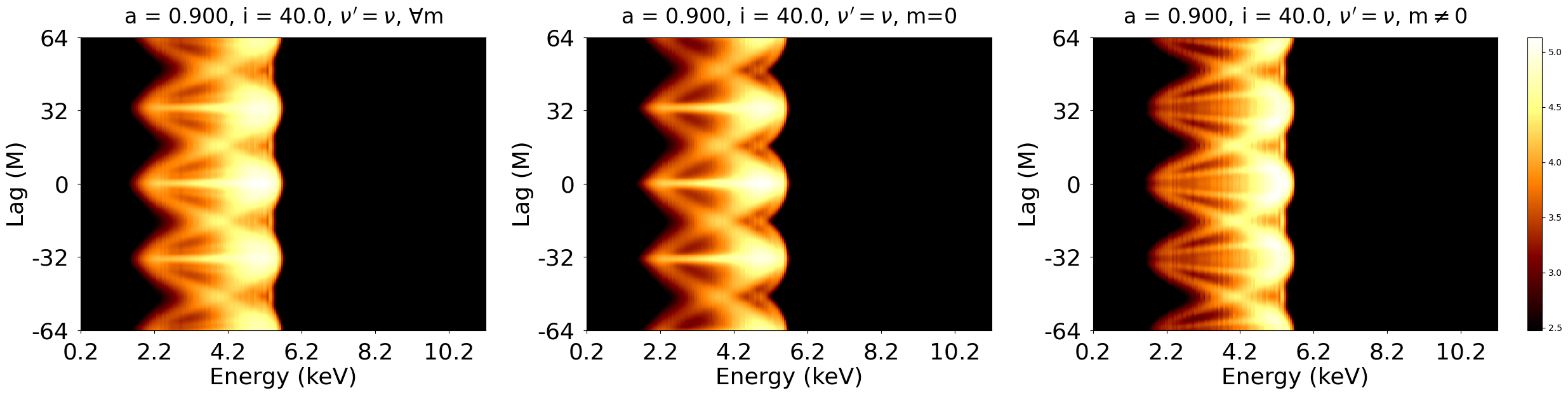}
    \includegraphics[width=1\linewidth]{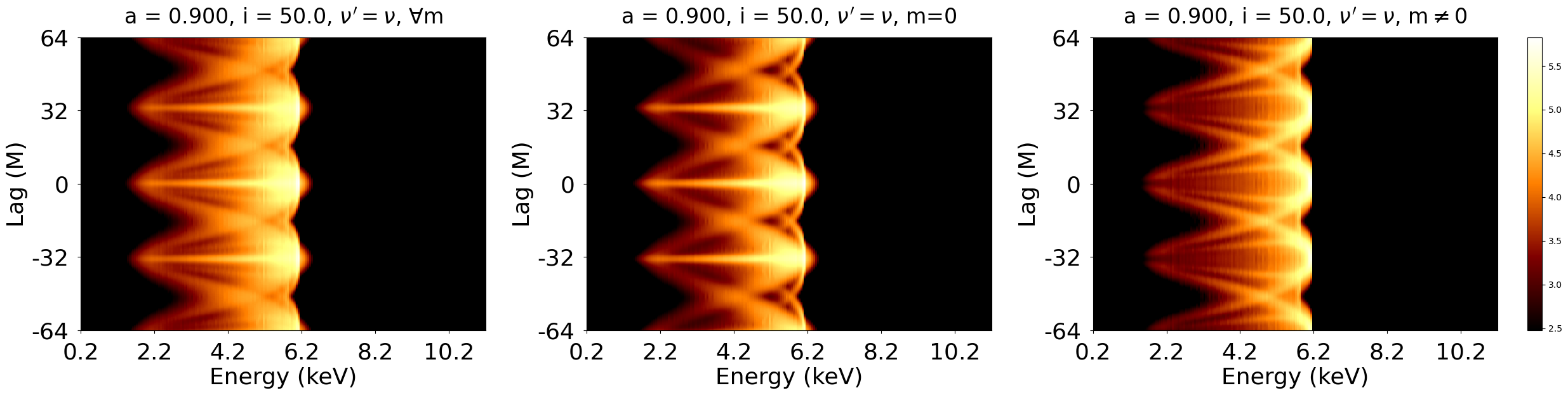}
    \includegraphics[width=1\linewidth]{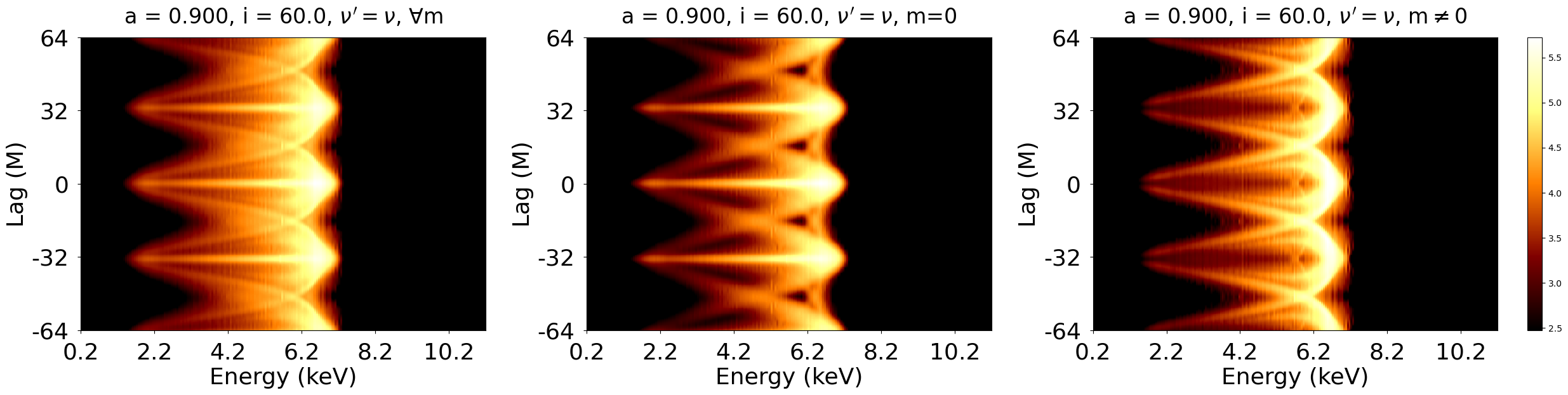}
    \includegraphics[width=1\linewidth]{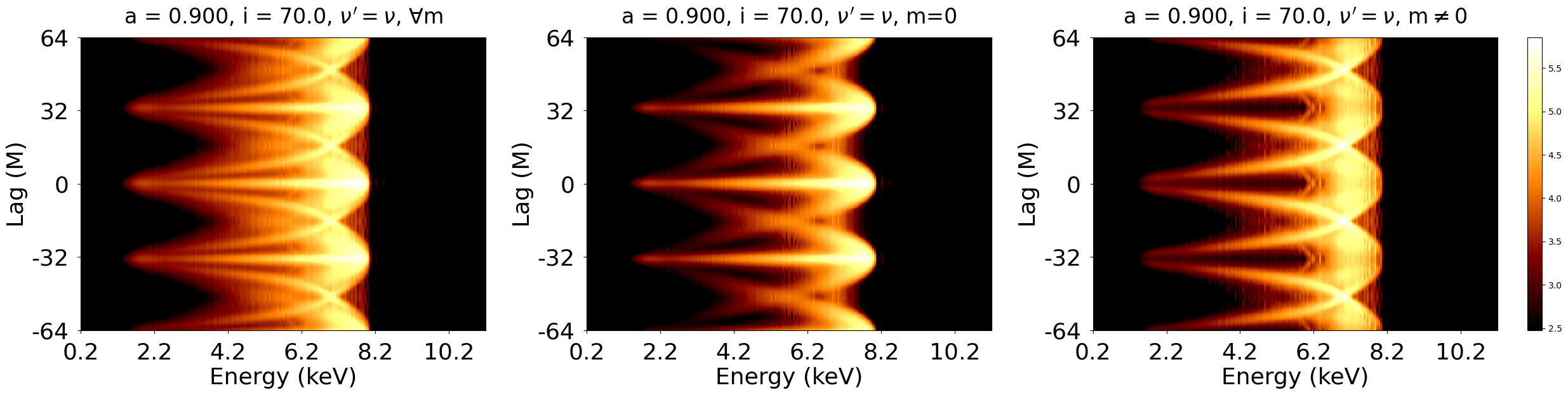}
    \caption{For a dimensionless black hole spin $a=0.9$, we plot here correlations (arbitrary units) of light curves of different energy (without mean substraction) considering a hotspot source revolving at the innermost stable circular orbit. Different rows correspond to different inclinations in the range 40$^\circ$-70$^\circ$.
    The first column represents the total correlation maps comprised of both lensing-dominated and direct emission-dominated photons. The second column is direct emission-dominated ($m=0$) and the third column represents the lensing-dominated ($m\neq0$) photons alone. We can clearly identify that the total image has contributions from the lensed signal. This is more evident at inclinations of 50 to 70$^{\circ}$.}
    \label{fig:Lenseffect}
\end{figure}

One may either choose to correlate the full light curves $f_\nu(t)$ with each other, or to correlate only the fluctuations $\Delta f_\nu(t) = f_\nu(t)-\langle{ f_\nu(t) \rangle}$, subtracting their mean values.
Fig. \ref{fig:Lenseffect} shows the correlation patterns generated with the full light curves, without mean subtraction.
The rows correspond to different inclination values of 40$^{\circ}$, 50$^{\circ}$, 60$^{\circ}$, and 70$^{\circ}$ respectively. The first column represents the total signal, which corresponds to that observed with a telescope. The second column represents only the direct emission-dominated correlations corresponding to $m=0$ and the third column portrays the lensing-dominated part, $m\neq0$. 
When different rows are compared, it is clear that the shape of the correlation pattern expands along the energy axis with an increase in inclination; this effect will be discussed below in \S~\ref{sec:hotinc}.

For clarity, we will consider the third row that represents an inclination of 60$^{\circ}$. In the left panel representing the total correlation, we can identify features that are contributed separately by the middle ($m=0$) and right ($m\neq0$) panels. Thus, in principle, an observation of such correlations from actual light curves could allow to distinguish lensing-dominated and direct emission-dominated contributions visually. This also implies that any method used to extract spin and inclination information from the correlation maps could harness these lensing features to do so. Note that we used here non-mean-subtracted light curves and in order to generate these plots, we have set a lower threshold value of 300 below which the plot is black in color. This corresponds to 50-60\,dB below the maximal correlation value, which brackets the relevant dynamic range.

In the following sections, we discuss mean-subtracted (fluctuation) light curves, and use a different colormap to show the amplitude of correlations, now ranging from negative (blue) to positive (red) values.
Extreme lensing features are equally present when correlating fluctuations. This may be seen in many of the correlation maps displayed in the paper (Figs. \ref{fig:Inceffect1}--\ref{fig:Mass}). Fig.~\ref{fig:Spineffect1}, in particular its three upper rows, shows an especially elegant example of how the correlations decompose into $m=0$ and $m\neq0$ contributions.

\subsection{Effect of source parameters on correlation maps}
\label{sec:hotinc}
\begin{figure}
    \centering
    \includegraphics[width=1\linewidth]{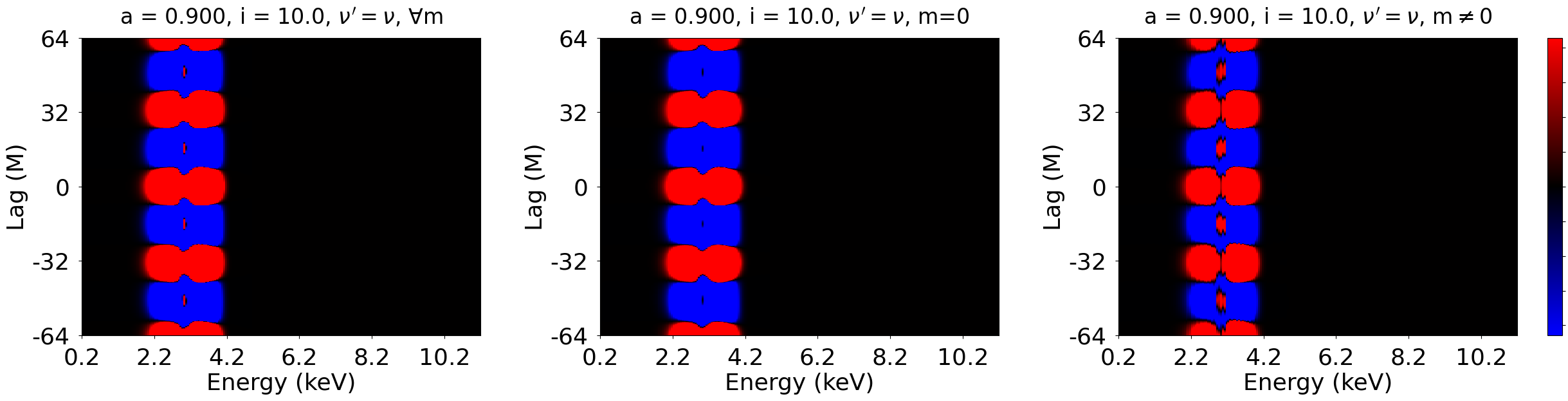}
    \includegraphics[width=1\linewidth]{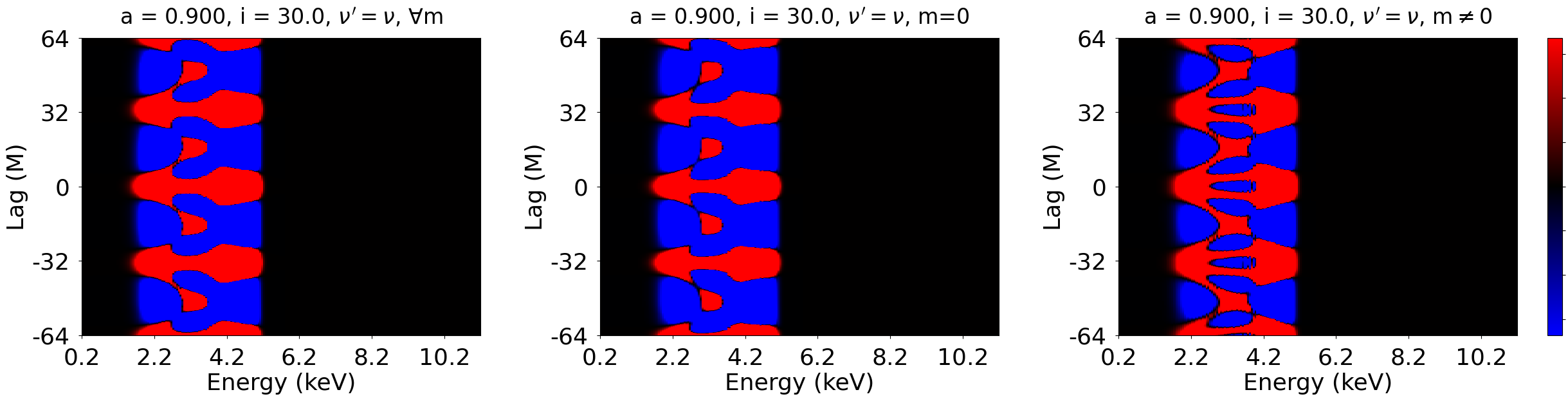}
    \includegraphics[width=1\linewidth]{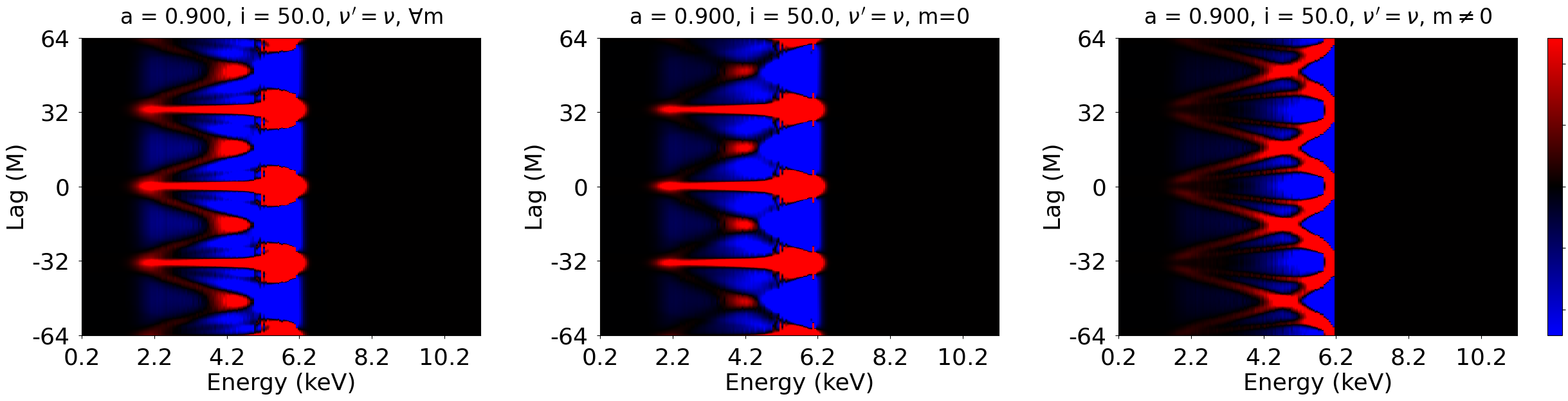}
    \includegraphics[width=1\linewidth]{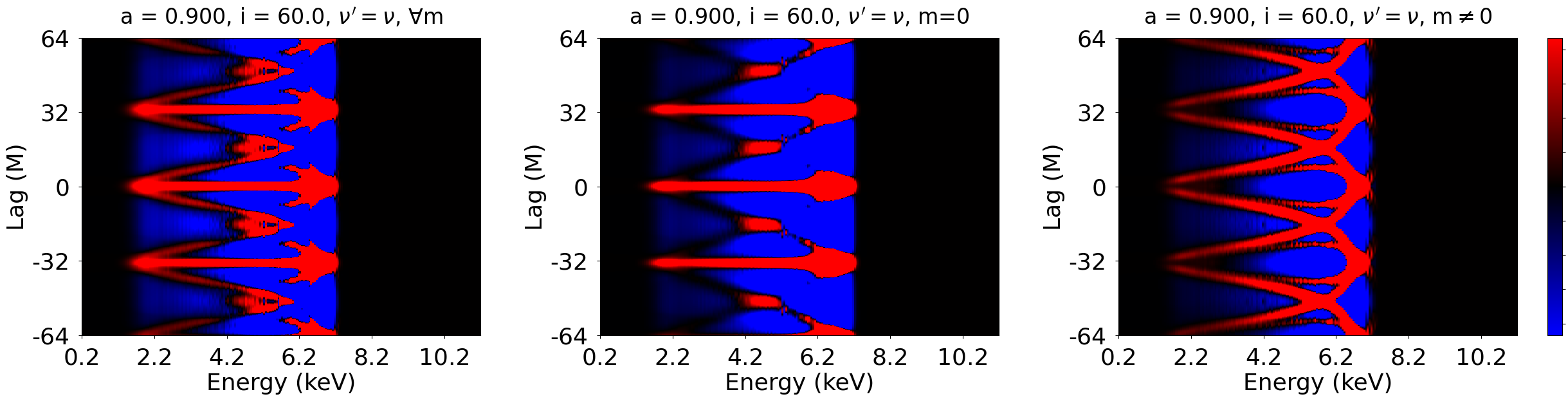}
    \includegraphics[width=1\linewidth]{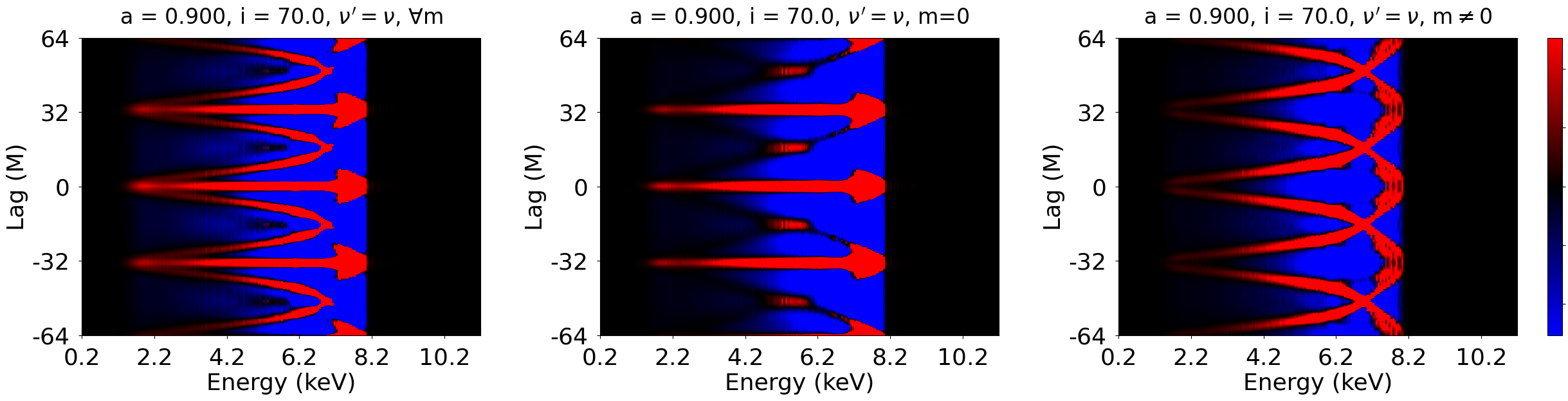}
    
    \caption{The dependence of correlation maps on inclination is depicted here for a fixed spin of 0.9. The inclination increases row wise. The columns represent $\forall$m (total), $m=0$ (direct emission-dominated) and m$\neq 0$ (lensing-dominated) respectively. We use arbitrary units and blue color indicates negative values, black indicates zero and positive values are represented in red. 
    }
    \label{fig:Inceffect1}
\end{figure}

\begin{figure}
    \centering
    \includegraphics[width=1\linewidth]{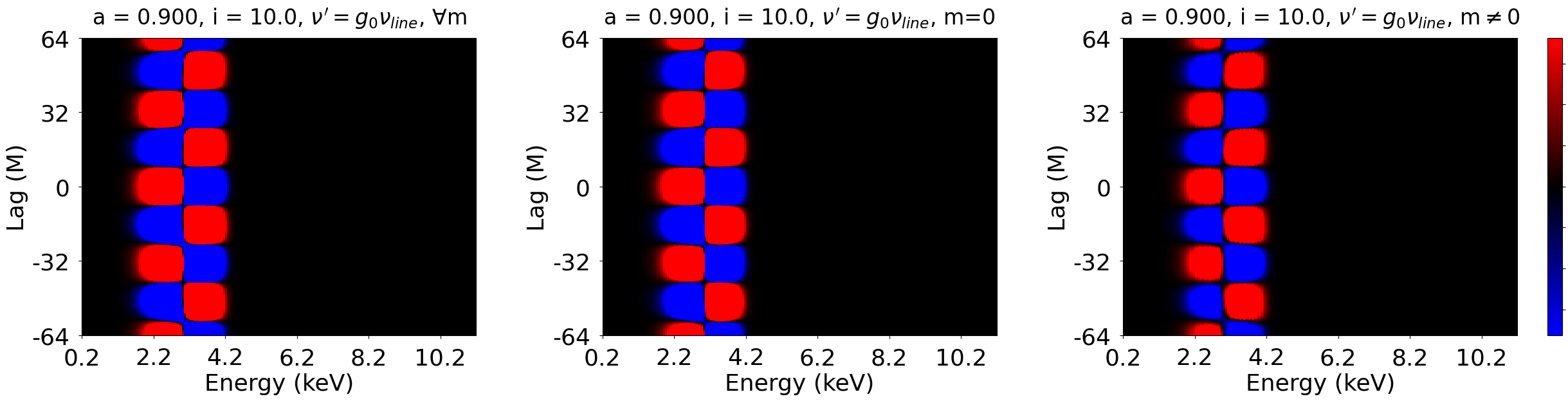}
    \includegraphics[width=1\linewidth]{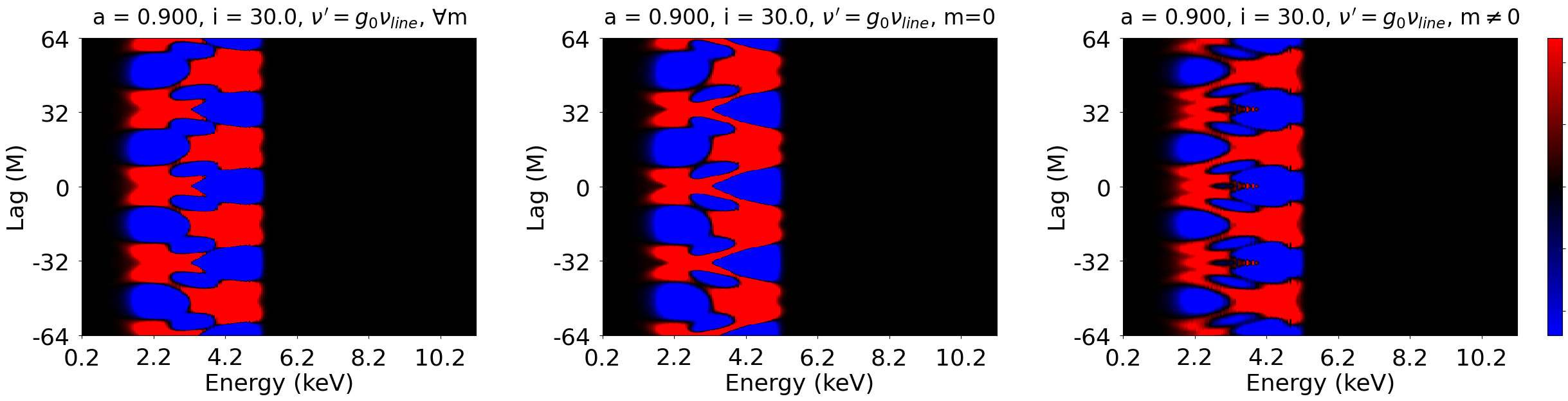}
    \includegraphics[width=1\linewidth]{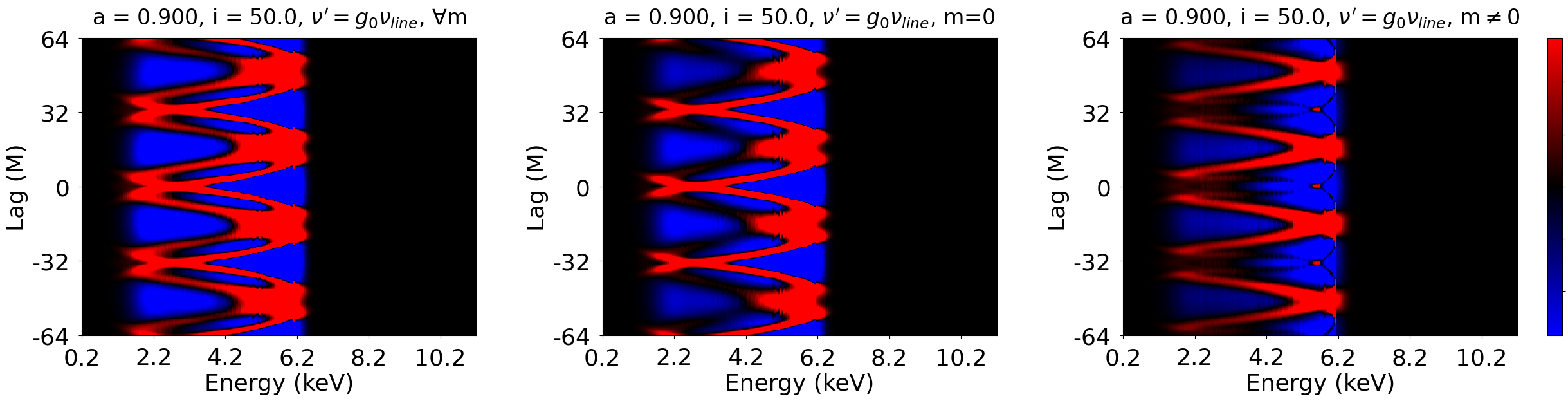}
    \includegraphics[width=1\linewidth]{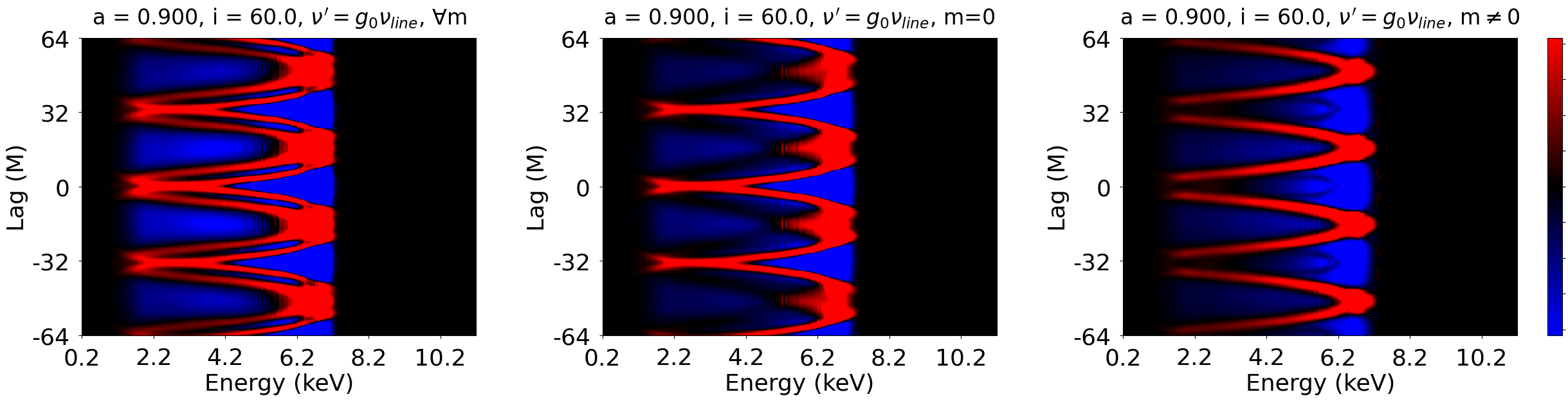}
    \includegraphics[width=1\linewidth]{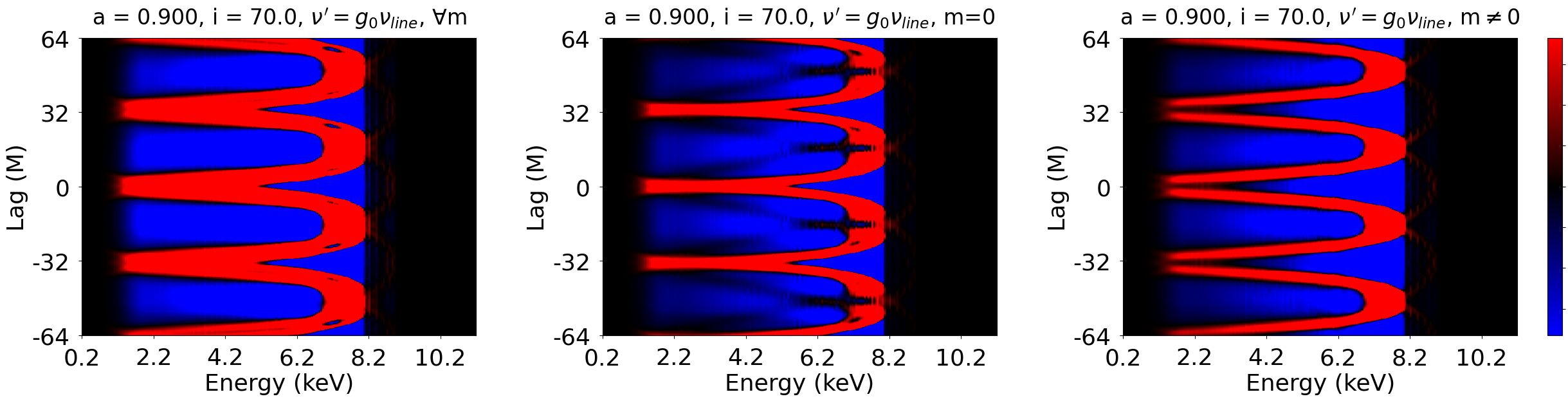}
    
    \caption{The effect of inclination on correlation patterns (in arbitrary units), visualized on the slice $\nu^{\prime}=g_0\nu_{line}$, for a (fixed) spin of 0.9. The inclination increases row wise. The columns represent $\forall$m (total), $m=0$ (direct-emission-dominated) and m$\neq 0$ (lensing-dominated) respectively.
    The pattern expands along the energy axis due to Doppler red/blueshift as inclination is increased. 
    }
    \label{fig:Inceffect2}
\end{figure}

Here we describe some heuristics of the dependence of correlation maps on source parameters. We first discuss the dependence on source inclination angle. This has to do with the Doppler effect due to hotspot motion along the line of sight, redshifting or blueshifting the observed signal, and in particular stretching the energy range of observed photons at higher inclinations \cite{Seward2010}.  
In Fig.~\ref{fig:Inceffect1} we show some examples of correlation maps computed for an orbiting hotspot at ISCO for a spin of 0.9, observed at different inclination angles ranging from 10$^\circ$ to 70$^\circ$.
Note that we have used mean-subtracted light curves to generate these figures.
It can be clearly seen that, as expected, when the source inclination increases, the correlation pattern expands along the energy axis.  In this figure, we once again present the total (all $m$) signal on the left panel, its direct emission-dominated ($m=0$) part in the middle panel, and the lensing-dominated part ($m\neq0$) on the right panel. 
The presence of lensing effects in the left panel is quite evident mostly at high inclinations such as 60$^{\circ}$ and 70$^{\circ}$. For the other slicing of the correlation matrix, with $\nu^{\prime}=g_0\nu_{line}$, we have presented the variations with inclination in Fig.~\ref{fig:Inceffect2}. With an increase in source inclination, we see an expansion of patterns along the energy axis which can be attributed to the Doppler boosts along the line of sight as the hotspot orbits the BH. 

\begin{figure}
    \centering
    \includegraphics[width=1\linewidth]{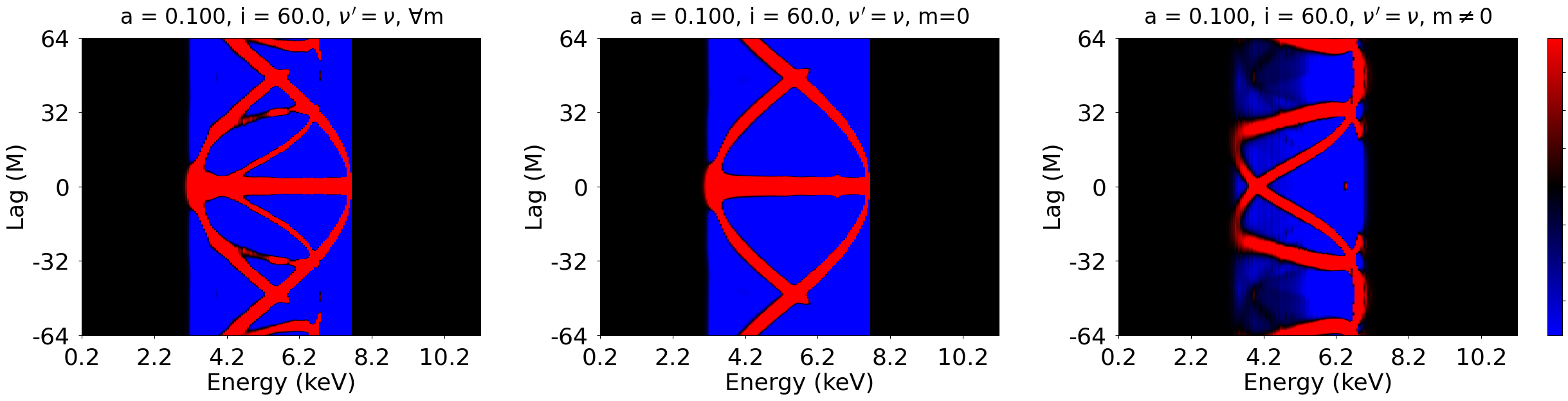}
    \includegraphics[width=1\linewidth]{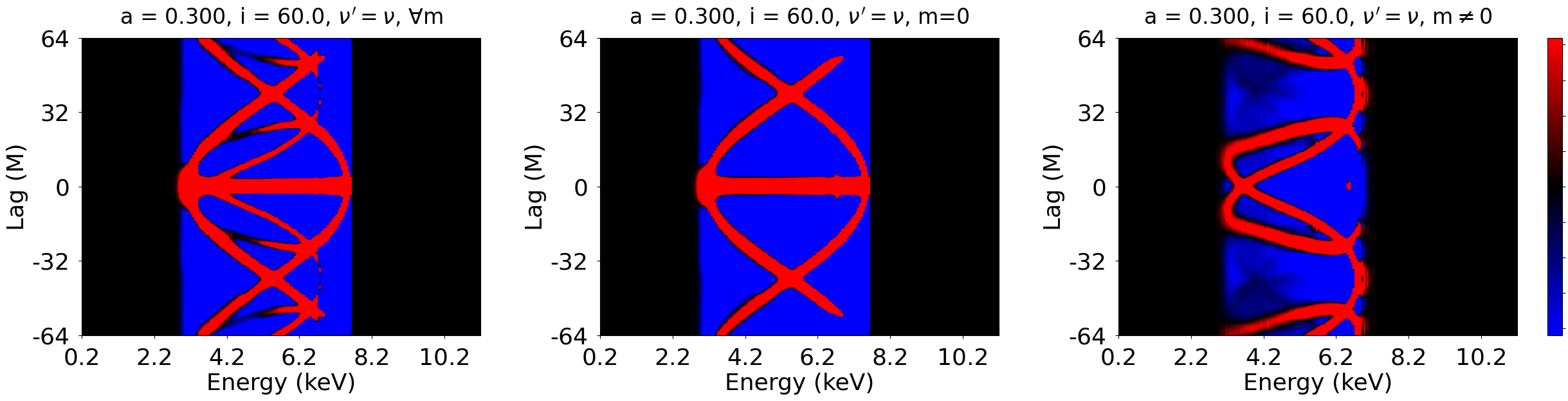}
    \includegraphics[width=1\linewidth]{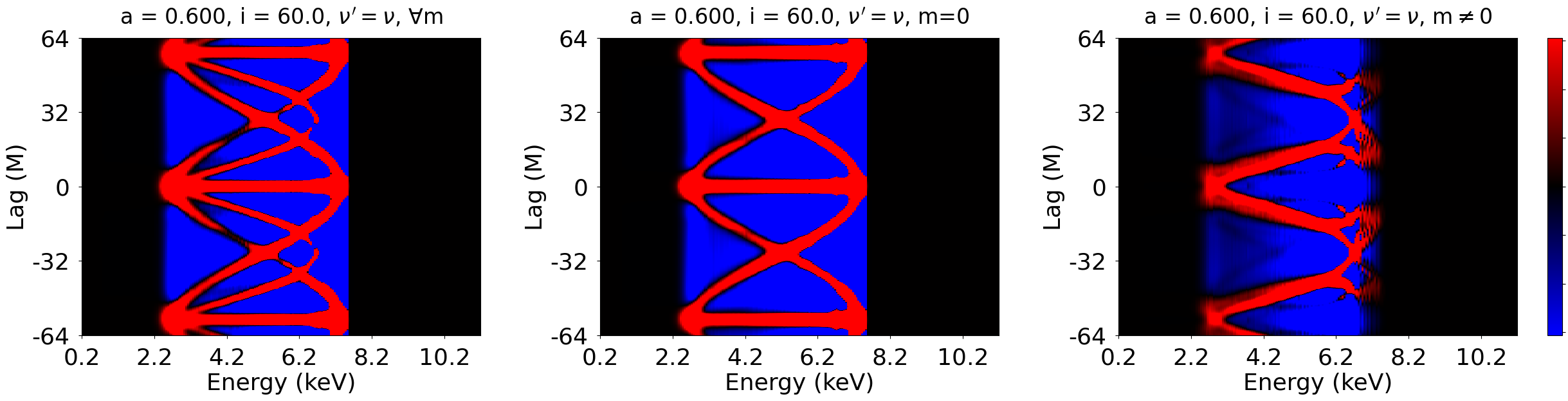}
    \includegraphics[width=1\linewidth]{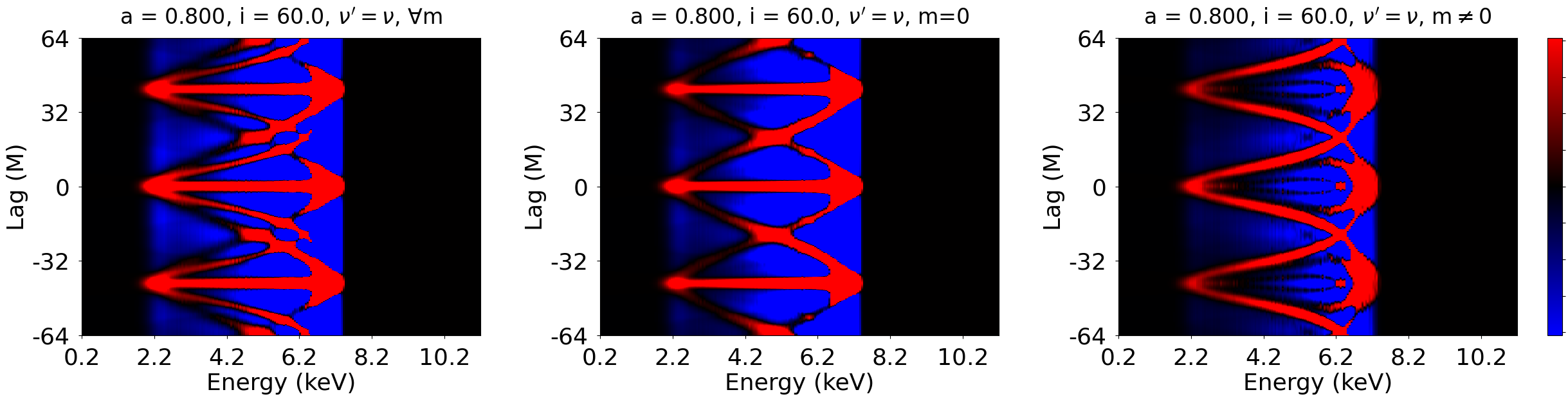}
    \includegraphics[width=1\linewidth]{a_0.9_i_60_C_nuprime_E_0.2-11.2keV_3in1_check-cdict-vmaxby10d.png}
    \caption{Dependence of correlation maps (in arbitrary units) on spin for a hotspot at the ISCO, viewed at a constant inclination of 60$^{\circ}$. The spin increases row wise. The columns represent $\forall$m (total), $m=0$ and m$\neq 0$ (lensing-dominated) respectively.
    We can observe that the pattern shrinks along the lag axis
    since the ISCO orbital period decreases with spin. }
    \label{fig:Spineffect1}
\end{figure}

\begin{figure}
    \centering
    \includegraphics[width=1\linewidth]{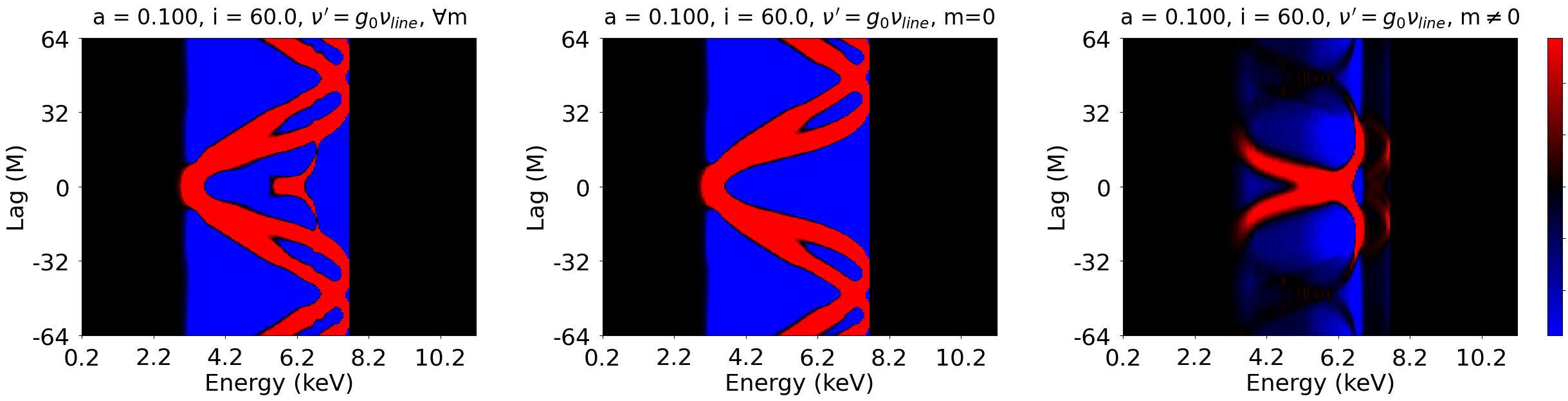}
    \includegraphics[width=1\linewidth]{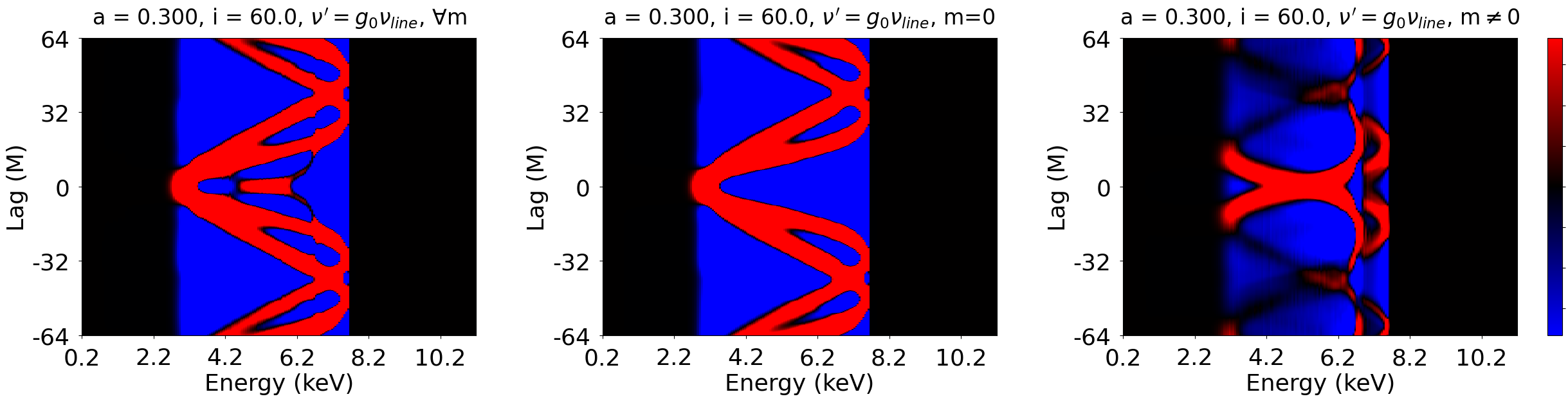}
    \includegraphics[width=1\linewidth]{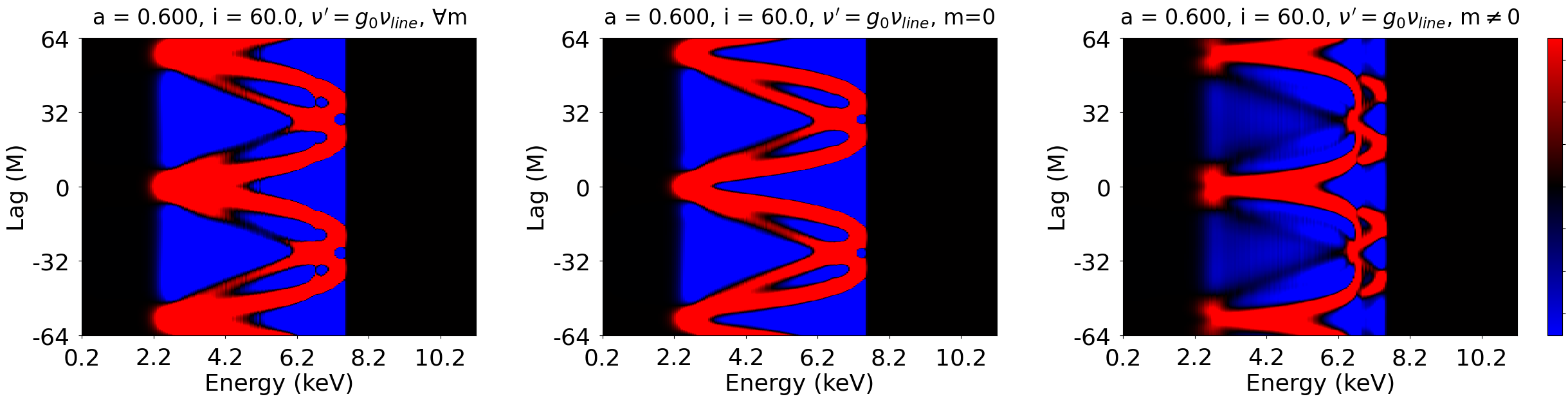}
    \includegraphics[width=1\linewidth]{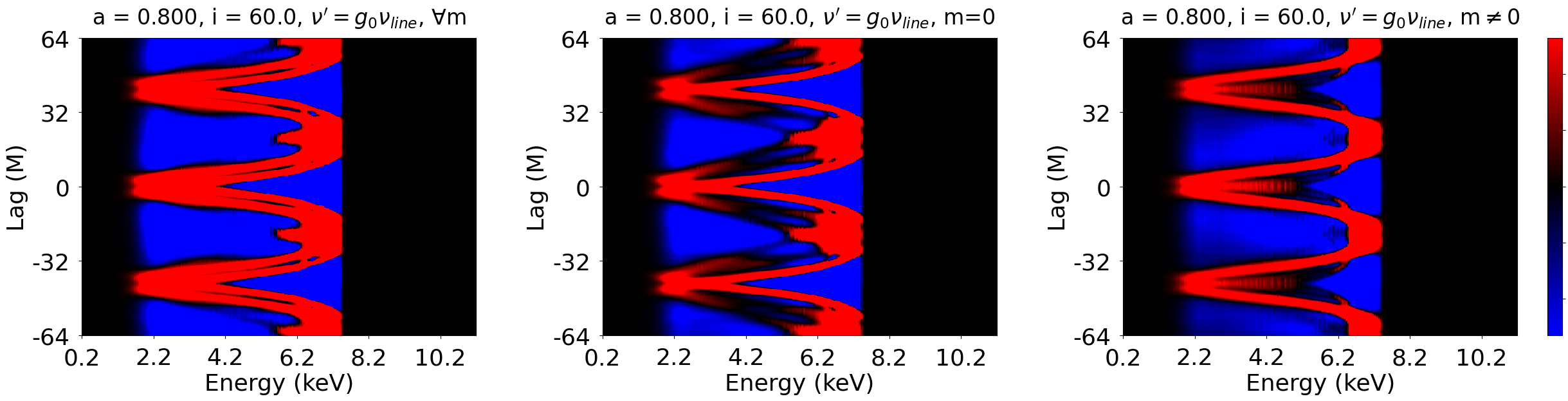}
    \includegraphics[width=1\linewidth]{a_0.9_i_60_C_nuline_E_0.2-11.2keV_3in1_check-cdict-vmaxby10d.png}
    \caption{The effect of spin on correlation maps for a hotspot at the ISCO. Here we show $\nu^{\prime}=g_0\nu_{line}$ slices, the inclination is fixed to 60$^{\circ}$, and the spin increases row wise. The columns represent $\forall$m (total), $m=0$ and m$\neq 0$ (lensing) respectively. The correlations are in arbitrary units and the blue indicates negative values while red indicates positive values.}
    \label{fig:Spineffect2}
\end{figure}

Any new insight into BH spin inference would be highly valuable from an observational perspective, and this motivates us to understand how BH spin affects spectrotemporal correlations. We begin for simplicity with a constrained model in which the hotspot is positioned at the ISCO.
Within this model, Fig.~\ref{fig:Spineffect1} shows how the increase in spin (row-wise) shrinks the correlation patterns along the lag axis. The spin values considered here are 0.1, 0.3, 0.6, 0.8, and 0.9. The pattern shrinkage has to do with the fact that the hotspot that moves along a geodesic at $r=r_s$ has a period $P=2\pi M\left((r_s/M)^{3/2}+a\right)$. If we moreover require $r_s=r_{\mathrm{isco}}$, the period at fixed BH mass becomes a rather strongly-varying function of the spin alone, and this period becomes shorter as the ISCO radius decreases with increasing spin. For the case of $\nu^{\prime}=g_0\nu_{line}$, we have presented the variations of the correlation maps with spin, again within the constrained model of an ISCO hotspot, in Fig.~\ref{fig:Spineffect2}. A similar evolution of patterns as in the previous case is visible here as well. It should be stressed that although we get strong spin dependence on correlation patterns in the ISCO-constrained model, 
this constraint is likely not a realistic one. From the expression for the period above, we see that if the orbiting radius $r_s$ of the hotspot assumes a different, say random value other than ISCO, then the impact of BH spin on the orbital period diminishes. Interestingly, in this more realistic case we find that extreme lensing signatures seem to provide a sensitive proxy of BH spin. We further discuss this point in \S \ref{sec:radhs}.

\begin{figure}
    \centering
    \includegraphics[scale=0.22]{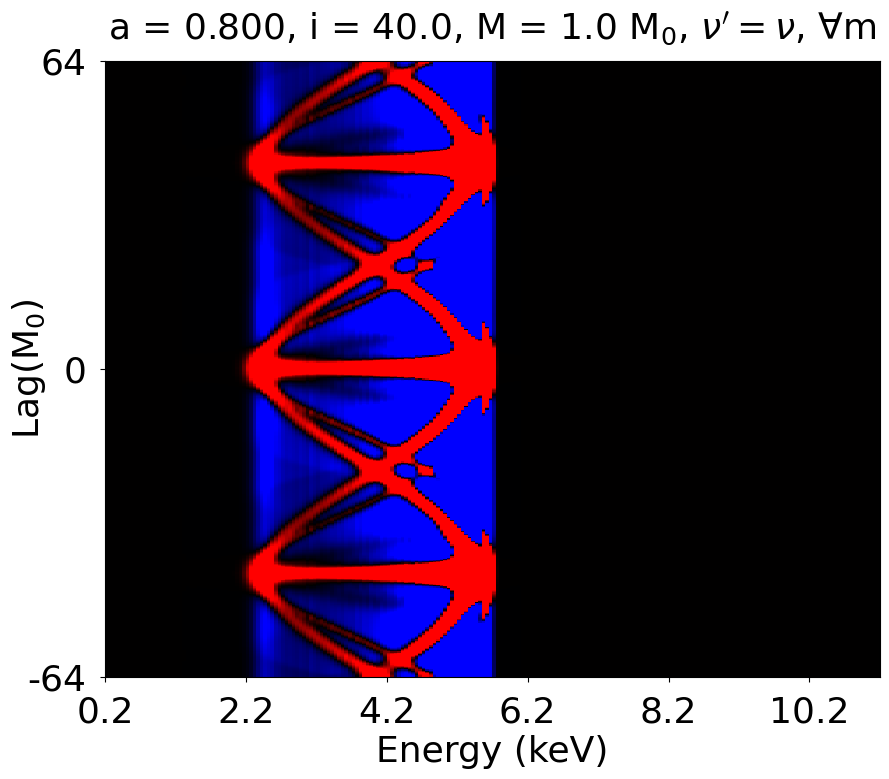}
    \includegraphics[scale=0.22]{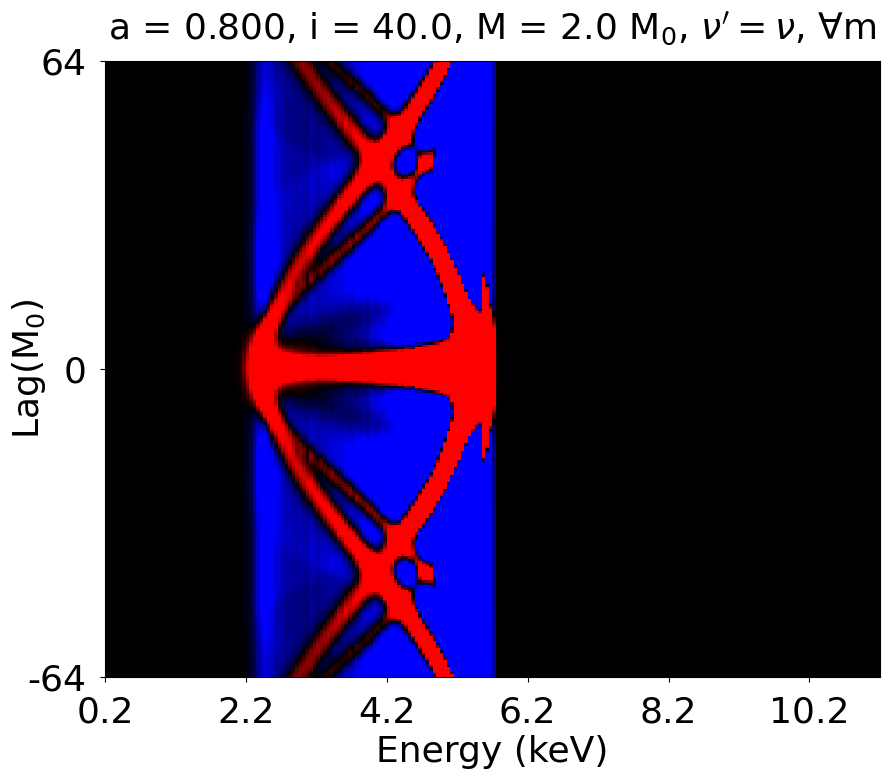}
    \includegraphics[scale=0.22]{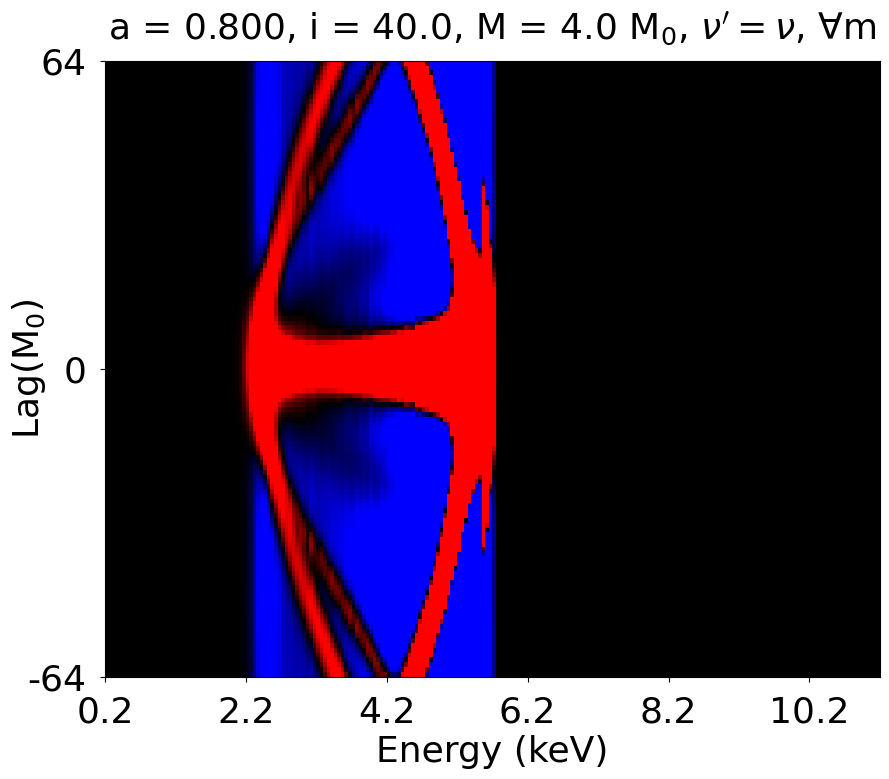}

    \caption{Dependence of the correlation patterns on the black hole mass. 
    As mass grows, the pattern expands along the lag axis. The image shows patterns for masses 1 $M_0$, 2 $M_0$, and 4 $M_0$, where $M_0$ is some reference mass.}
    \label{fig:Mass}
\end{figure}

Finally, we point out that all time lags in the problem (arising from orbital periods as well as extreme lensing) scale linearly with BH mass, when all other dimensionless parameters are kept fixed. This relation is portrayed in Fig.~\ref{fig:Mass}, which presents correlation maps for sources of different masses, all with dimensionless spin $a=0.8$ and an inclination of 40$^{\circ}$. We can clearly see that the effect of mass variation is to `stretch' correlation patterns along the lag axis.

In the next section, we will describe how we constructed a machine learning tool that leverages these heuristics and allows to infer BH parameters from correlation maps within our hotspot model.

\section{Black hole parameters with Machine Learning}
\label{sec:ML}
\subsection{2D Convolutional Neural Network}
\begin{figure}
    \centering
    \includegraphics[width=1\linewidth]{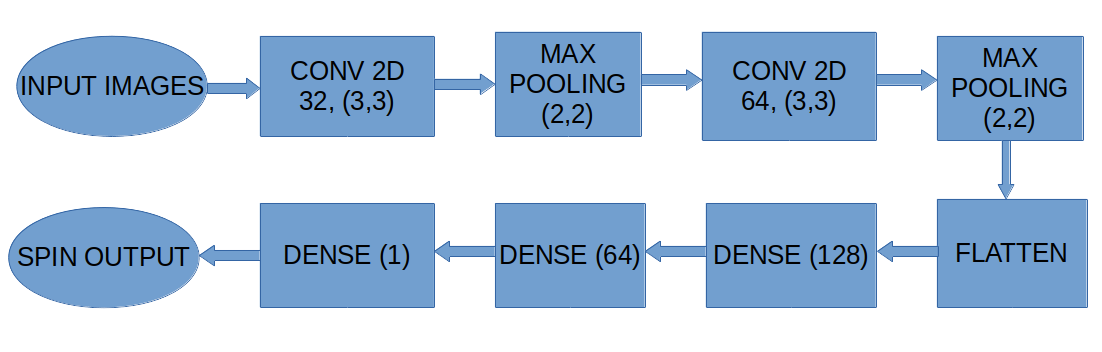}
    \caption{The neural network for spin inference involves two convolutional layers followed by corresponding maxpooling layers before the three fully connected dense layers. The specifications of each layer are provided in the flowchart. We train an identical network for the inclination parameter as well.}
    \label{fig:2Dflowchart}
\end{figure}

We generated data containing 280 2D correlation maps, each corresponding to random values of spin (between 0.02 and 0.998), inclination (between 1$^{\circ}$ and 89$^{\circ}$), and hotspot size (between 0.1 M to 0.5 M). Each correlation map is a slice taken from a 3D correlation matrix by applying the criterion $\nu^{\prime}=\nu$. This correlation array in the time range of -64 M to +64 M is extracted from a highly time-resolved (0.25 M) ray tracing output. As it is computationally expensive and memory-exhaustive, we used a basic set of 280 images that we will later increase to 470, as reported in \S \ref{sec:radhs}. We used these images to train a 2D CNN using the \texttt{tensorflow} package in Python.
Before training the model, the images were preprocessed by resizing them to (224 x 224) and all pixel values were normalized by dividing them by 255 as the pixel values are in the range (0,255).  
We observed that the models are more accurate if we use separate CNNs to detect spin and inclination. The flowchart for the model is shown in Fig. \ref{fig:2Dflowchart}.

For the case of spin estimation, we used a CNN starting with a 2D convolution layer comprising of 32 filters each of size (3 x 3). Following it we introduced a max-pooling layer having a kernel size of (2 x 2). Then we flatten the outputs and feed them to a fully connected (Dense) layer of 128 neurons having a rectified linear unit (ReLU) as the activation function. This is then processed through another \texttt{Dense} layer of 64 neurons. The results are then fed to the final layer having one neuron that will predict the value of spin. 

We compile the model using `adam' optimizer and the loss is computed using the mean squared error (mse) value. 
We train the model for spin parameter in 20 epochs having a validation split of 15 percent data.
We have kept aside 15 \% of the available data for testing. The division of data into training and test sets were done using \texttt{train\_test\_split} feature in \texttt{scikit-learn} package by setting a random seed value of 42. The results showed a root mean squared error (rmse) of 0.11. The model estimates the spin values from 0.15 to 0.9 with a reasonably good rmse of 0.06. In the spin range of $0.15<a<0.95$, the accuracy is 0.09. The error in spin estimates exceeds the net rmse of 0.11 outside the above spin range. These results are for the cases where we know the hotspot sizes (0.5 M). If we do not know the hotspot's size, we can estimate spin within an error of $\pm$ 0.13 and inclination within an error bar of 4.7$^{\circ}$. This is achieved by training the deep learning model with random hotspot half-widths in the range 0.1 M to 0.5 M. Then we introduce a Gaussian noise of standard deviation equal to 10\% the value of the correlation matrix. This has given a model with an error in spin of 0.13 while the error in inclination is 4.68$^{\circ}$, implying that the noise does not change the estimates significantly. However, if we consider hotspots that can be located anywhere around the BH rather than fixed to the ISCO, the dependence of these correlation patterns on source spin will become less straightforward. In such a case, we expect that the model for inclination inference will not be affected much, though the model for spin estimation may become less accurate. We test this in the following subsection.

\subsubsection{Introducing random hotspot radius}
\label{sec:radhs}
To make the model more realistic, instead of fixing the hotspot at $r_s=r_{\rm{isco}}$, we consider hotspots randomly located in the range $r_{\rm{isco}} \leq r_s \leq 10M$, imagining a situation in which the relevant hotspots are generated only in the close vicinity of the BH. 
We also highlight that we have increased the number of basic correlation images used from 280 to 470 by introducing sources with random combinations of spin, inclination, hotspot width, and hotspot orbital radius (between ISCO and 10 M). Remarkably, a CNN model trained to estimate the radius of the hotspot performs quite well and can predict its radial position within an error bar of 0.44 M.

\begin{table}[]
    \centering
    \begin{tabular}{|c|c|c|c|c|c|c|}
    \hline
        &\multicolumn{3}{|l|}{With 3290 images}&\multicolumn{3}{|l|}{ With 6110 images} \\              \hline
        Parameter & $\forall$ m & m=0 & m$\neq$0 & $\forall$ m & m=0 & m$\neq$0 \\
        \hline
        Inclination &  1.59$^{\circ}$ & 1.41$^{\circ}$ & 2.55$^{\circ}$ &  1.48$^{\circ}$ & 1.01$^{\circ}$ & 1.58$^{\circ}$\\
        Spin & 0.21 & 0.19 & 0.12 & 0.13 & 0.12 & 0.08\\
        Mass factor (k) & 0.26 & 0.28 & 0.36 & 0.19 & 0.19 & 0.27\\
        Hotspot location (M) & 0.29 & 0.22 & 0.18 & 0.23 & 0.16 & 0.09\\
       \hline
    \end{tabular}
    \caption{The root mean squared error in source parameter estimates with a 2D CNN trained using 3290  correlation maps for the case $\nu^{\prime}=\nu$ is presented in the first three columns. The rmse for parameters when using a CNN model trained on a larger set of 6110 images is presented in the last three columns. We considered masses from 1M to 4M in steps of 0.5 M in the first case and 0.25 M in the second case.}
    \label{tab:2DCNN}
\end{table}

\subsubsection{Incorporating the mass parameter}
In \S \ref{sec:hotinc}, we mentioned how the change in the mass parameter affects the correlation patterns. Implementing this idea, we generated more correlation maps for masses $kM_0$, where $M_0$ is a reference mass and $k$ was taken to vary from 1 to 4 in steps of size 0.25. Thus, we increased our sample size from 470 images to 6110 (470 $\times$ 13) images. Training a neural network with these images, we could estimate inclination with an accuracy of 1.48$^{\circ}$, spin with an accuracy of 0.13, orbital radius within an error of 0.23 M and mass factor $k$ with an accuracy of 0.19. These results are presented in Table \ref{tab:2DCNN}. 

If we consider only images corresponding to $m=0$ (direct emission-dominated correlations), the spin estimate remains comparable to the above case with an error of 0.12. Interestingly, if we use only lensing-dominated ($m \ne 0$) correlations, the spin estimate improves so that the error on the spin becomes 0.08. This indicates that the lensing features are strongly influenced by the spin of the source and as a result if we could detect extreme lensing signatures, they could be used for estimating BH spin with better accuracy. However, even without lensing correlations, a CNN based on $m=0$ images also proves to be efficient in estimating the spin within the confines of our simple toy model.

\subsection{3D Convolutional Neural Network}
\begin{figure}
    \centering
    \includegraphics[width=1\linewidth]{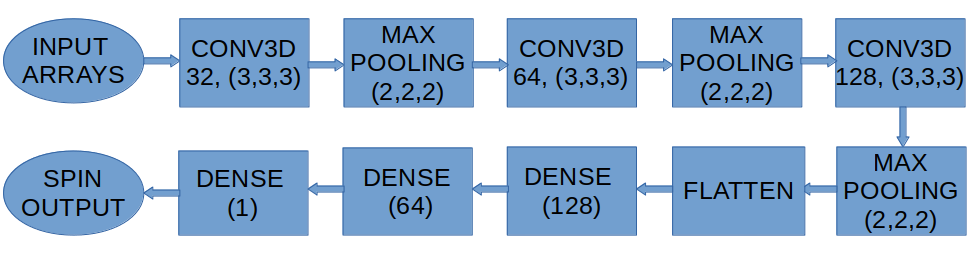} 
    \caption{The neural network for spin inference involves three convolutional layers followed by corresponding maxpooling layers before the three fully connected dense layers. The specifications of each layer are provided in the flowchart. We train an identical network for the inclination parameter as well.}
    \label{fig:3Dflowchart}
\end{figure}
Here, we wish to utilize the full 3D information encoded in the correlation function. Using the 280 three-dimensional arrays of size 220x220x512, we trained a 3D CNN with 3 layers of convolution and \texttt{maxpooling} followed by three fully connected dense layers for spin estimation as shown in Figure \ref{fig:3Dflowchart}. For the case of the inclination parameter, we used an identical network. The filters used were of size 3x3x3 and the maxpooling used a kernel of size 2x2x2. The data arrays were normalized for each time instance, which improved the training of the model. The root mean square error for training was 0.06. The mean absolute error for training (mae), validation and testing was 0.04, 0.06, and 0.04, respectively. This implies that the model gives fairly accurate results in the estimation of the spin. However, to make it more realistic, when we trained the model with random hotspots between 0.1 M and 0.5 M, the results were spin estimates with an rmse of 0.13 and inclination estimates with an rmse of 6.4$^{\circ}$.

\subsection{Model comparison}
The accuracy in spin parameter estimates did not improve significantly as we increased the size of the basic training data. This shows the limitation in spin estimates at extreme values of spin from correlation data, as we further discuss below. Note that the error bars in inclination estimates using a 3D CNN based on 3D correlation matrices were larger than those for the 2D CNN model. 
Further we trained 2D CNN models based on different cases like $\nu' = \nu - 1$keV/h and also with 2 keV offset. The results show comparable error bars as with the case of $\nu'=\nu$ and hence they have no preference over the results included here.

To crosscheck how well the 2D CNN model estimates spin in specific ranges, we trained the CNN with only spins ranging from 0.5 to 0.9 (a subset having 2210 images). The results show that this model can constrain spin with an rmse of 0.04, which is better than the error of 0.13 obtained with the full set of correlation maps for all m using 6110 images. So, the CNN models work better in the spin range of 0.5-0.9 which does not consider the more uncertain estimates at very high and low spin values. However, the net result of 0.13 rmse in spin implies that the model is reliable for any source spin value. By introducing a random gaussian noise within 50\% the standard deviation of each correlation matrix, the error in spin estimates increased to 0.20 compared to 0.13 in the noise-free case. For the inclination estimate the error went from 1.5$^{\circ}$ to 1.7$^{\circ}$ with the addition of Gaussian noise.

\section{Observational Considerations}
\label{sec:ForObs}
In our previous paper on spectrotemporal correlations \cite{Hadar2023}, we have estimated the time resolution as well as exposure time required for a signal-to-ratio of 1 for both BH binaries and supermassive BHs. The important features of the telescope required for this method are its large collecting area, an operating energy range of 0.2-12 keV, and a good spectral resolution. Upcoming missions such as NewAthena \cite{Athena_10FIRST_2013} significantly reduce the required integration time, improving the chance of detecting lensing features. This is mainly due to NewAthena's photon collecting area of 1.4 m$^2$, that is at least 10 times larger than any of the currently operational X-ray telescopes. The proposed spatial resolution of 5 arcseconds in the NewAthena mission would suffice for our case because we deal only with X-ray emissions, and these are produced only from the inner regions of the source. However, a realistic detectability calculation with SNR estimates considering instrumental noise for upcoming missions is beyond the scope of this paper.

Another aspect to consider is the source that can be observed to use our method based on X-ray correlations. 
Even though the method can work with all ranges of spin and inclination, a source with inclination between 45-75$^{\circ}$ is preferred in order to easily differentiate between the lensing-dominated and direct emission-dominated features in the correlation maps. This is because the correlation patterns expand with inclination and the constituent features become more visible at larger inclinations. GRS 1915+105 with an inclination of 66$\pm2^{\circ}$ \cite{Fender1999,Zd2014} is possibly a good candidate. Observations during the quiescent phase of BH binaries such as GRS 1915+105 are more appropriate to detect subtle features 
due to extreme lensing, as the signal reaching the observer will be less obscured by variability and accretion processes that will be dominant during an outbursting/active phase.

\section{Discussion and Conclusion}
\label{sec:disc}
In this paper, we have explored correlation patterns in BH spectroscopic time series within a simple source model, investigating how they may be used to detect lensing effects and infer BH parameters. 
Our work is motivated by the prospect of providing a novel method for probing extreme lensing effects by BHs, even when they are spatially unresolved. A successful detection of such effects, if it materializes, would be commensurate with a measurement of the photon ring and could eventually be used to draw information about numerous targets which are otherwise out of reach.
Such data could pave the way for testing current paradigms of BH spin evolution with cosmic time
and its dependence on BH mass and the orientation of the accretion flow. For instance, Ref.~\cite{Dotti2013} argues that spin evolution can be erratic and highly influenced by the disc angular momentum for low-mass BHs, as opposed to very massive sources, which are more likely to gradually spin down in the absence of a preferential direction of accretion. In another example, BHs of extremely high spin are believed to have accumulated mass due to prolonged nonchaotic accretion, as both numerical relativity simulations \cite{Berti2008} and current LIGO-VIRGO-KAGRA data \cite{LVK2023} seem to indicate that high-spin BHs are rarely the result of mergers. Cosmological simulations with \texttt{NewHorizon} and \texttt{Galactica} suggest that in the late phase of evolution, the spin of BHs tends to align with the angular momentum of its host galaxy \cite{Peirani2024}. A technique to estimate the rotation speed of the source from movies made out of consecutive EHT images has been proposed by \cite{Conroy2023}.

Technically, our main tool was the Kerr ray-tracing code AART \cite{Alejandro2023}, with which we compute `movies', including frequency information. By spatially integrating over snapshots, we generate light curves of the specific flux $f_\nu(t)$ and compute correlations between them, such that the output is a 3D matrix $C_{\nu \nu'}(T)$ depending on energy, energy, and lag. For convenience, in this paper we often focus on two special slices of the full correlation matrix, defined by $\nu^{\prime} = \Tilde{g}_0\nu_{line}$ and $\nu^{\prime} = \nu$, generating 2D correlation maps.

Our results for the correlation maps allow us to explore the imprints of extreme lensing embedded within them. Considering different combinations of source parameters---BH mass, spin, and inclination, and hotspot radial position and size---allows an exploration of parameter dependence and an improved heuristic understanding of STACs beyond the universal regime, i.e., for low-$n$ contributions. In Fig.~\ref{fig:Lenseffect} we show that the total correlation pattern includes both lensing-dominated and direct emission-dominated contributions. Here, the lensing part is especially prominent at large inclinations of around 50$^{\circ}$ - 70$^{\circ}$. This has to do with the fact that at higher source inclinations the correlation patterns expand along the energy axis because of Doppler effects along the line of sight. The same principle is demonstrated in Figs.~\ref{fig:Inceffect1}~and~\ref{fig:Inceffect2}, where we study correlations of mean-subtracted (fluctuation) light curves.
In Figs.~\ref{fig:Spineffect1}~and~\ref{fig:Spineffect2} we present the spin dependence under the assumption that the hotspot is located at the ISCO. Correlation patterns are straightforwardly sensitive to spin under this assumption, since the hotspot's orbital period significantly varies with spin. 
However, fixing the hotspot at the ISCO seems much too restrictive, and allowing a random orbital radii for hotspots should significantly increase our model's realism. In that case, the correlation patterns must still depend on spin, albeit in a more intricate manner, namely through lensing effects; especially extreme lensing. 
This expectation proved to be correct, as we discuss below.

In \S \ref{sec:ML}, we present the use of deep learning and machine learning algorithms to estimate the parameters of a BH from its correlation maps. We generated correlation patterns for various combinations of spin, inclination, hotspot width (random uniform), and random hotspot orbital radius (between ISCO and $10M$) and trained CNN models to estimate the source parameters. The network performed well in estimating these parameters with high accuracy. CNNs that were trained on 2D and 3D correlation matrices helped us generate models that could detect inclination within an error bar of $\approx 1.5^{\circ}$, dimensionsless spin within an error bar of $\approx 0.13$ and mass with an accuracy of $\Delta M/M \approx0.2$. These models are robust to varying hotspot radius (under the assumption of a geodesic circular orbit) as they were trained with random hotspot radii, and could predict the orbital radius of the hotspot with a precision of $\approx0.23 M$. See Table~\ref{tab:2DCNN} for more details on error estimates in parameter inference.
In order to see the effect of noise on the correlation patterns, we have introduced random Gaussian noise within a range of 50\% the standard deviation of each correlation matrix. The results showed an increase in error from 0.13 to 0.20 for the spin parameter estimate from the total correlation pattern. In the case of low (nearly zero) or high (near-extremal) spin, the correlation patterns we found 
were usually (for most choices of inclination, hotspot location, and hotspot size) insensitive to slight variations in spin. 
However, it would be interesting to carry out a separate and more detailed
analysis of correlations for the very high spin case; such an analysis is beyond the scope of this work.
It is interesting to note a property of correlation-based parameter inference that arises from our study. If we use only lensing-dominated contributions to train the CNN and infer parameters, we get significantly more sensitivity to spin (error of $\approx0.08$) than when using the total correlations. 
This split is, of course, artificial, but it shows the effectiveness of lensing-dominated correlations in the inference of spacetime properties. This could have been expected: as discussed in \S~\ref{sec:introduction}, lensing-dominated quantities---just like the photon ring in BH images---probe the spacetime geometry more directly than direct emission-dominated quantities. See further discussion in \S \ref{sec:radhs}.

Our results suggest a number of natural generalizations.
One may consider a more general hotspot: noncircular, nongeodesic, and/or nonequatorial. 
It would also be interesting to provide a simple geometric-statistical model of variable coronal emission---that is, without modeling its physics but only the emissivity profile and fluctuations---and study the corresponding STACs. 
Thus, analyzing correlation patterns of such observations can help us understand the physical scenario at play in the vicinity of a BH.
Another interesting scenario to study, which could improve on model realism, would be to assume that the source is a stochastic accretion disk with some prescribed statistical properties, as has been done in \cite{Hadar2021,Hadar2023} analytically. Numerical modelling of such a disk---not of the physics but only of the statistical properties and average spatial emissivity profile---has been discussed in \cite{Lee2021,Alejandro2023} in the context of BH imaging. We could of course use hints from accretion disk simulations \cite{Pech2008} and extend the current work by using multiple randomly fluctuating hotspots or flares at various locations around the black hole. Orbiting hotspots would contribute correlation enhancements at the respective orbital periods, and there would be additional contributions due to correlation amongst different hotspots in the images. 
It would be interesting to use the insights acquired in this work in order to numerically compute photon ring autocorrelations \cite{Hadar2021}, the cousin of the present STAC observable, which is relevant for BH interferometric imaging.
It could also be interesting to generalize the observables under study, for example, to include the effects of polarization.

Finally, it would be interesting to conduct a search for STACs, of the type studied in the present paper, in observed spectroscopic time series. Of course, if absorption can be neglected, the patterns that are expected in observation are the combined ones, i.e., of the type of the first column of figures \ref{fig:Lenseffect}, \ref{fig:Inceffect1}, \ref{fig:Inceffect2}, \ref{fig:Spineffect1}, and \ref{fig:Spineffect2}. Data from proposed missions like NewAthena \cite{Athena_10FIRST_2013} which will have an order-of-magnitude improvement in collecting area over current missions such as XMM-Newton, Neil Gehrel's Swift Observatory, XRISM, and NICER could enable us to identify lensing signatures in correlation maps. However, as we have shown in Table \ref{tab:2DCNN}, even without detecting lensing signals with current telescopes (that is, from images where m$\neq$0 is not visible), the resulting correlation maps may be used to estimate black hole parameters with good accuracy, within our model. Further, with the help of simulations and computations such as the ones carried out in this paper, and by incorporating more complicated disc emissivity profiles, we could be able to use models based on CNNs to accurately estimate the source spin, mass, and inclination.

\section*{Acknowledgements}

We are grateful to Alejandro C\'ardenas-Avenda\~no and Ilan Shimshoni for valuable discussions. We acknowledge financial support by the Data Science Research Center (DSRC) at the University of Haifa. S. Harikesh is supported by a Bloom postdoctoral fellowship. S. Hadar is supported in part by a grant from the Israeli Science Foundation (ISF 2047/23). The research of D. Chelouche is partially supported by grants from the German Science Foundation (DFG HA3555-14/1, CH71-34-3) and the Israeli Science Foundation (ISF 2398/19, 1650/23). The computations presented in this work were performed on the Hive computer cluster at the University of Haifa, which is partly funded by ISF grant 2155/15.

\section*{Data Availability}
The data that support the findings of this article are openly available \cite{Data_zenodo}.
\bibliography{references.bib}
\bibliographystyle{utphys}

\end{document}